\documentclass{aa}  

\usepackage{graphicx}
\usepackage{xcolor}
\usepackage{txfonts}
\begin{document} 
\newcommand{\Rer}{\ensuremath{R_{\oplus}}}
\newcommand{\Mer}{\ensuremath{M_{\oplus}}}
\definecolor{dk}{rgb}{0.,0.,0.}

   \title{The mass-radius relation of intermediate-mass planets outlined by hydrodynamic escape and thermal evolution}

   \subtitle{}

   \author{D. Kubyshkina
          \inst{1}
          \and
          L. Fossati \inst{1} 
          }

   \institute{Space Research Institute, Austrian Academy of Sciences,
              Schmiedlstrasse 6, A-8042 Graz, Austria\\
              \email{daria.kubyshkina@oeaw.ac.at}
             }

   \date{Received, 2022; accepted, 202?}

  \abstract
   {Exoplanets in the mass range between Earth and Saturn show a large radius, and thus density, spread for a given mass.} 
   {We aim at understanding to which extent the observed radius spread is affected by the specific planetary parameters at formation and by planetary atmospheric evolution, respectively.} 
   {We employ planetary evolution modeling to reproduce the mass--radius (MR) distribution of the 198 so far detected planets with mass and radius measured to the $\leq$45\% and $\leq$15\% level, respectively, and less massive than 108\,$M_{\oplus}$. We simultaneously account for atmospheric escape, based on the results of hydrodynamic simulations, and thermal evolution, based on planetary structure evolution models. Since the high-energy stellar radiation affects atmospheric evolution, we account for the entire range of possible stellar rotation evolution histories.
   To set the planetary parameters at formation, we use analytical approximations based on formation models. Finally, we build a grid of synthetic planets with parameters reflecting those of the observed distribution.}
   {The predicted radius spread reproduces well the observed MR distribution, except for two distinct groups of outliers ($\approx$10\% of the population). The first group consists of very close-in Saturn-mass planets with Jupiter-like radii for which our modeling underpredicts the radius likely because it lacks additional (internal) heating similar to that responsible for inflation in hot Jupiters. The second group consists of warm ($\sim$400--800\,K) sub-Neptunes, which should host massive primordial hydrogen-dominated atmospheres, but instead present high densities indicative of small gaseous envelopes ($<$1--2\%). This suggests that their formation, internal structure, and evolution is different from that of atmospheric evolution through escape of hydrogen-dominated envelopes accreted onto rocky cores. On average the observed characteristics of low-mass planets ($\leq$10--15\,$M_{\oplus}$) strongly depend on the impact of atmospheric escape, and thus of the evolution of the host star's activity level, while primordial parameters are less relevant. Instead, for more massive planets, the parameters at formation play the dominant role in shaping the final MR distribution. In general, the intrinsic spread in the evolution of the activity of the host stars can explain just about a quarter of the observed radius spread.}
   {}

   \keywords{hydrodynamics --
                Planets and satellites: atmospheres --
                Planets and satellites: fundamental parameters -- Planets and satellites: formation --
                Planets and satellites: physical evolution
               }
   \titlerunning{Mass-radius relation of intermediate-mass planets}
   \maketitle
%
\section{Introduction}\label{sec::intro}
The population of low- and intermediate-mass exoplanets displays a wide radius, and thus density, spread \citep[e.g.][]{weiss2014,hatzes2015,ulmer2019}. These planets, which consist of super-Earths and sub-Neptunes (hereafter altogether called sub-Neptunes, unless specified otherwise), are typically low-mass planets with radii smaller than $\sim$4\,\Rer\ and comprise about 60\% of all exoplanets known to date. In many cases, the mass of these planets is unknown due to the difficulty of measuring the small radial velocity signals they typically produce on the host stars, particularly if faint as for most of the stars hosting sub-Neptunes detected by the Kepler satellite. This results in just about 11\% of all detected sub-Neptunes having a measured mass. Therefore, most works dedicated to study the parameter distribution of these planets, such as the radius gap \citep[i.e. the lack of planets with radii between about 1.5 and 2\,\Rer; e.g.][]{fulton2017}, rely on \textcolor{dk}{synthetic underlying mass functions, whose shape is uncertain and that typically ignore planets more massive than $\sim$20\,\Mer} \citep[see, e.g.][]{wu2019,gupta_schlichting2019,modirrousta2020,rogers_owen2021}.

The small number of sub-Neptunes with an accurately measured mass has forced exoplanet population studies to focus mainly on the analysis of the radius vs period (or orbital separation) plane, with the eventual addition of the stellar mass as a further parameter \citep[e.g.][]{owen_wu2017,jin2018}. However, space- and ground-based facilities (e.g. TESS, CHEOPS, HARPS, HARPS-N, HIRES) have started focusing on detecting planets orbiting bright stars, and thus enabling one to accurately measure planetary masses and radii. By adding the planetary mass among the observables, population studies have gained one more fundamental parameter to constrain planetary formation and evolution models. Therefore, in this work we aim to reproduce the exoplanet mass-radius (MR) distribution employing basic results of planetary formation models and detailed planetary atmospheric evolution models.

The large radius spread observed for sub-Neptunes is commonly believed to be the direct consequence of the diversity of planetary atmospheres. In particular, planets are considered to be composed by rocky (silicates and heavier metals) cores surrounded by extended hydrogen-dominated (for sub-Neptunes) or compact secondary (for super-Earths) atmospheres, where the radius gap represents the boundary between these two regimes \citep[e.g.][]{owen_wu2017,jin2018}. In this context, the most prominent factor shaping the radius distribution is atmospheric mass loss removing primordially accreted atmospheres affecting different planets in different ways and over different time-scales \citep[e.g.][]{owen_wu2013,owen_wu2017,ginzburg2018,gupta_schlichting2019}. This explanation finds observational support in the shape of the radius distribution and in the position of the radius gap varying with planetary orbital separation, system age, and mass and metallicity of the host stars, in agreement with the predictions of atmospheric mass-loss models \citep[e.g.][]{fulton2018,david2021gap,sandoval2021,Petigura2022}.

However, the internal structure of sub-Neptunes can be significantly more diverse than that of a rocky core surrounded by a gaseous atmosphere. For example, these planets might hold significant fractions of ices or liquid water \citep[e.g.][]{rogers_seager2010,dorn2017structure}. Therefore, \citet{zeng2021gap_waterEOS} suggested that the radii of planets hotter than 900\,K and masses below 20\,\Mer\ can be reproduced assuming ice-dominated compositions without significant gaseous envelopes, and thus the entire MR distribution of sub-Neptunes can be outlined by purely rocky planets, gas-poor water worlds, and gas-rich water worlds, where the former two represent the two peaks in radius distribution.

Some studies indicated that assuming realistic protoplanetary disk parameters enables one to reproduce the radius distribution of sub-Neptunes purely on the basis of planetary formation processes. \citet{venturini2020} showed that the two peaks of the radius distribution can be reproduced by metal-rich rocky cores (super-Earths) and icy cores (sub-Neptunes). However, in this case the further addition of an atmospheric layer leads to overestimating the fraction of large sub-Neptunes. 
In general, it is expected that close-in planets with masses below $\sim$20--30\,\Mer\ are strongly affected by atmospheric escape, while this process does not play a significant role in the evolution of more massive planets. However, the observed MR diagram presents a significant spread in radius for a given mass also for higher masses up to $\sim$0.3\,$M_{\rm jup}$, that is $\sim$100\,\Mer. Therefore, one could expect that for low-mass planets the radius spread can be caused by atmospheric escape, while for massive planets the radius spread has a primordial origin. For example, it has been shown that for close-in planets with masses $\leq$5--10\,\Mer\ atmospheric escape leads them onto the same MR relation within a few tens of Myr, independently of the initial atmospheric parameters \citep[e.g.][]{kubyshkina2020mesa,kubyshkina2021mesa}. However, \citet{hallat2022subsaturns} showed that even short-period sub-Saturn-mass planets ($\sim$60--100\,\Mer) can experience significant atmospheric mass loss throughout their lifetime, which makes less clear the separation of the relative impact of primordial parameters and atmospheric escape on planetary evolution. Therefore, it appears that the boundary between the escape- and formation-dominated regimes in terms of impact on planetary evolution depends on the system parameters and in particular on planetary mass, orbital separation, and stellar properties.

In this study, we aim at quantifying the impact of atmospheric mass loss onto the observed MR distribution of planets across a wide range of masses. To this end, we employ an atmospheric evolution framework combining realistic hydrodynamic escape and planetary thermal evolution \citep{kubyshkina2021mesa,kubyshkina2022evo} applying it to a grid of planets in the 1--110\,\Mer\ range on various orbits corresponding to planetary equilibrium temperatures ($T_{\rm eq}$) not lower than 500\,K ($\sim$2.5--75\,days). 
In particular, we aim at outlining the role played by the initial atmospheric parameters at formation and by atmospheric escape processes as a function of system parameters. Simultaneously, we aim at identifying which is the fraction of the radius spread that can be explained by the range of possible rotation history of the host stars, which we use as a proxy for stellar activity, and thus high-energy (X-ray and extreme ultraviolet; XUV) emission finally incident on a planet.

The paper is organized as follows. In Section~\ref{sec::model}, we describe the modeling approach, which includes the framework employed for tracking the evolution of planetary atmospheres (Section~\ref{sec::model::model}), the set of model planets used to outline the observed population (Section~\ref{sec::model::planets}), the considered sets of initial conditions at the time of the protoplanetary disk dispersal (Section~\ref{sec::model::initials}), and the adopted stellar evolution models (Section~\ref{sec::model::mors}). In Section~\ref{sec::observations}, we describe the criteria used to select the detected planets on which to compare the modeling results, briefly discussing the distribution of planetary and stellar parameters within the observational data set. We compare the modeling predictions with the observations and discuss the main features of the theoretical population in Section~\ref{sec::compare}. We discuss the results and draw the conclusions in Section~\ref{sec::discussion} and Section~\ref{sec::conclusions}, respectively.
   
\section{Modeling approach}\label{sec::model}
\subsection{Physical model}\label{sec::model::model}
To model the evolution of planetary atmospheres, we employ an upgraded version of the framework developed by \citet{kubyshkina2020mesa} and \citet{kubyshkina2021mesa}. It combines atmospheric structure and thermal evolution modeling performed using MESA \citep[Modules for Experiments in Stellar Astrophysics][]{paxton2011,paxton2013,paxton2018} with atmospheric mass-loss rates based on detailed hydrodynamic modeling \citep{kubyshkina2018grid,kubyshkina_fossati2021}.

First, we use MESA to obtain the atmospheric structure of a planet with a given set of parameters. These parameters include total planetary mass, mass or density of the planetary core (i.e. the solid and inert internal part of the planet composed of silicates and metals, defining the lower boundary of the simulation domain), atmospheric mass fraction, initial entropy (temperature) of the core, and bolometric irradiation from the host star setting up the temperature at the upper boundary. The latter is defined by the stellar properties, the planetary orbital separation, and the age of protoplanetary disk dispersal. 

For planets with masses below 25\,\Mer, we set up the planetary model at the time of protoplanetary disk dispersal following the algorithm described by \citet{chen_rogers2016}, and thus produce an initial model representing a coreless ball of gas with a mass of 0.1 Jupiter masses ($\sim 31.8$\,\Mer). Then, we (1) add the core of the desired mass and density without changing the total mass of the planet, (2) reduce the atmosphere to the desired mass fraction, (3) implement an artificial luminosity (or let the planet cool down) to reach the desired initial core entropy, and (4) relax the stellar irradiation to set up the upper boundary conditions. Although we applied it, the penultimate step is not strictly necessary, because the particular value of the initial entropy of sub-Neptunes plays a role in the evolution only in the first few tens of megayears and does not impact the evolution over Gyrs timescale \citep{owen2020entropies,kubyshkina2021mesa}. We specifically verified this statement for the planets considered in this work. For numerical reasons, for setting up the initial model of the heavier planets, we consider a 100\,\Mer\ planet with a 20\,\Mer\ core and swap steps (1) and (2).

Once steps (1) to (4) are completed, the planetary model is ready to initiate atmospheric evolution, which begins at the time of protoplanetary disk dispersal. The exact lifetime of the disk, which varies within the 2--10\,Myr range \citep[e.g.][]{mamajek2009}, has a minor effect on the results. Therefore, we set this time arbitrarily at 5\,Myr. At this time, we switch on atmospheric mass loss based on interpolation within the grid of hydrodynamic upper atmosphere models published by \citet{kubyshkina_fossati2021}. The grid covers a wide range of planetary and stellar parameters and is based on one-dimensional hydrodynamic models accounting for atmospheric escape due to XUV irradiation and internal thermal energy of the planet \citep{kubyshkina2018grid}. Both atmospheric $T_{\rm eq}$ and stellar XUV emission vary with time following the evolution of the host star described in Section~\ref{sec::model::mors}.
\subsection{Modeled planets}\label{sec::model::planets}
We split the considered 1--108.6\,\Mer\ planetary mass range into 20 distinct initial planetary masses. To minimise interpolation errors in the calculation of the atmospheric mass-loss rates, we chose the specific masses so that they coincide with the nodes of the grid of hydrodynamic upper atmospheres models. For each planetary mass, we further consider that each planet orbits a star of 1\,$M_{\odot}$ at a range of orbital separations, namely 0.03, 0.05, 0.1, 0.2, and 0.35\,AU. The chosen orbital separation range encompasses that of the vast majority of the detected planets with well measured masses and radii, and extending to longer separations would not add information when comparing model predictions with observations. Furthermore, at the longest orbital separation \textcolor{dk}{$T_{\rm eq}$ corresponds to 512\,K}, which is close to the lower $T_{\rm eq}$ boundary of 300\,K of the grid of upper atmosphere models. 
The inner edge of the orbital separation range has been chosen in such a way to lie close to the inner boundary of the grid of upper atmosphere models, which is driven by the upper $T_{\rm eq}$ boundary of 2000\,K. \textcolor{dk}{As we consider only one stellar host mass of 1\,$M_{\odot}$, the orbital separations of our model planets correspond to specific equilibrium temperatures, as given in Table~\ref{tab::sma_teq}. In the following, when comparing the model results with observations, we consider the temperature values rather than the orbital separations, because it enables one to more easily find correspondences with real planets orbiting stars different than 1\,$M_{\odot}$ \citep{kubyshkina2021mesa}.}
\begin{table}[]
    \centering
    \textcolor{dk}{
    \caption{Correspondence between equilibrium temperature and orbital separation considered in the models.}
    \label{tab::sma_teq}
    \begin{tabular}{c|c}
    \hline\hline
      a [AU] &  $T_{\rm eq}$ [K] \\
         \hline
       0.03  & 1700 \\
       0.05  & 1350 \\
       0.1   & 956  \\
       0.2   & 680  \\
       0.35  & 512  \\
    \end{tabular}
    }    
\end{table}
\subsection{Initial conditions}\label{sec::model::initials}
Given a set of initial parameters, our framework tracks the evolution of a hydrogen-dominated planetary atmosphere through time. However, it does not include a mechanism allowing one to define self-consistently the initial parameters, such as atmospheric mass fraction or temperature of a newborn planet. This requires employing detailed planetary formation and/or atmospheric accretion models. Therefore, we describe here our approach of setting up the initial state of an evolving planet.
\begin{figure}
   \centering
   \includegraphics[width=\hsize]{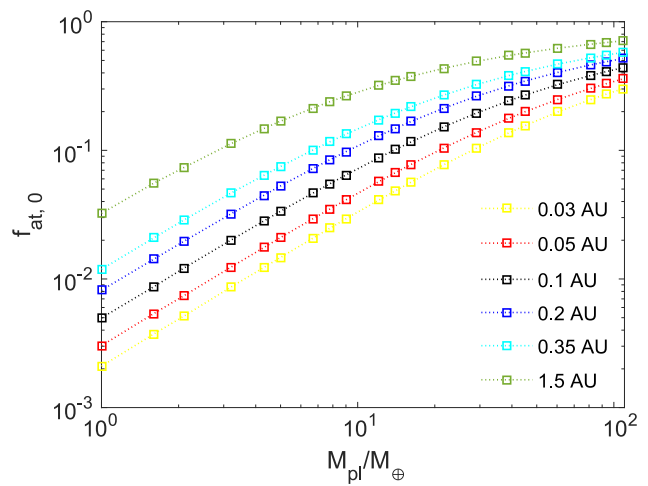} 
   \caption{Initial atmospheric mass fractions as a function of initial planetary masses for planets located at orbital separations of 0.03, 0.05, 0.1, 0.2, and 0.35\,AU according to the ``basic setup'' (Equation~(\ref{eq::fat_mordasini2020}); purple to cyan lines) and to the ``migrated'' scenarios (green line) defined in the text.}
   \label{fig::initials}%
\end{figure}

We do not aim to reconstruct in detail the formation mechanisms that shaped the primordial population of exoplanets across a wide range of masses, but rather to define to which extent the present-day population could be shaped by atmospheric mass loss or by the initial atmospheric mass fraction. In practice, we aim at identifying in which regions of the parameter space the initial conditions matter for determining the final state of evolving planets and where their imprint will they be erased by escape \citep{kubyshkina2020mesa}. %

Therefore, in the present study we do not employ full-scale formation models. Instead, to set up the initial parameters of the atmospheric evolution models, we use the analytical approximation of \citet{mordasini2020}, which is based on sophisticated population synthesis models by \citep{mordasini2014b}
\begin{equation}\label{eq::fat_mordasini2020}
    \frac{M_{\rm atm,0}}{M_{\rm core}} = 0.005\,\left(\frac{M_{\rm core}}{M_{\oplus}}\right)^{1.23}\,\left(\frac{a}{0.1\,{\rm AU}}\right)^{0.72}\,,
\end{equation}
where $M_{\rm atm,0}$ and $M_{\rm core}$ are the masses of the primordial planetary atmosphere and of the planetary core, respectively, and $a$ is the semi-major axis. The underlying model accounts for the accretion of planetesimals, orbital migration within the disc, and evolution of the protoplanetary disc. It assumes solar composition of H/He envelopes, and opacities from \citep[][grain-free]{freedman2014} and \citep[][grain opacities]{bell_lin1994}. The latter opacities were improved by \citet{mordasini2014a} by calibrating the gas accretion timescales with those found with the detailed model of \citet{Movshovitz2010} for grain dynamics.

\textcolor{dk}{Equation}~(\ref{eq::fat_mordasini2020}) predicts the mean atmospheric mass that is accreted by a core of a given mass at a specific orbital separation around a one solar mass star. We remark that the actual atmospheric masses can spread significantly (especially for the thinnest atmospheres) around this average value, depending on the specific parameters of the formation model not included explicitly in the equation above. 
Figure~\ref{fig::initials} shows the predictions of this approximation for the considered planetary masses and orbital separations \textcolor{dk}{in terms of initial atmospheric mass fraction $f_{\rm at,0} = M_{\rm atm,0}/(M_{\rm core} + M_{\rm atm,0})$}. Hereafter, we refer to these initial parameters as the ``basic setup'' scenario.

We remark that the distribution of planetary parameters employed to construct the models that led to Equation~(\ref{eq::fat_mordasini2020}) is not entirely compatible with that of the synthetic planets we consider in this work (Section~\ref{sec::model::planets}). In particular, the in-situ formation of sub-Neptunes at the two innermost orbital separations (i.e. 0.03 and 0.05\,AU) is unlikely, because of protoplanetary disc truncation \citep[e.g.][]{lee_chiang2017}. Therefore, many planets at such short orbital separations have likely formed further out and migrated inwards, which can then lead to a wider spread in initial (i.e. at the moment of protoplanetary disk dispersal) planetary parameters. The same holds true for planets with masses above the sub-Neptune range, namely above about 20--30\,\Mer. 

To account for planets that migrated inwards (potentially, from beyond the snow line), and thus having accreted more substantial envelopes, we considered planets with initial atmospheric mass fractions predicted by Equation~(\ref{eq::fat_mordasini2020}) for an orbital separation of 1.5\,AU (see Figure~\ref{fig::initials}). \textcolor{dk}{This choice does not have a significant impact on the results.} 
Therefore, \textcolor{dk}{to identify the maximum radius planets can have as a function of mass and temperature, we} further examine the evolution of these migrated planets in the considered range of orbital separations (0.03--0.35\,AU). Hereafter, we refer to these initial parameters as the ``migrated'' scenario.

According to the MESA simulations, the initial entropy of the planets described above varies in the 7--9.3\,${\rm k_{\rm b}/baryon}$ range, which corresponds roughly to 0.01--1000 times the present-day luminosity of Jupiter. Therefore, the planets are strongly inflated at the time of protoplanetary disk dispersal, which leads for some modeled planets to have large initial radii exceeding 10\,\Rer, and thus technically dropping out of the applicability range of the atmospheric escape grid. Therefore, we extrapolated the grid results, which we found being valid up to 15--20\,\Rer\ for massive planets and orbital separations beyond 0.1\,AU (where planets have the largest initial envelopes and typically experience low to moderate atmospheric mass loss). However, to verify that this has no impact on the final results, we re-simulated the evolution of planets with radii larger than 10\,\Rer\ considering smaller initial entropies, obtaining converging evolutionary tracks already after $\sim$100\,Myr. This approach is valid, because the initial entropy of a planet is important mostly during the first tens of Myr and has a negligible effect on planetary evolution over Gyr timescales \citep{owen2020entropies,kubyshkina2021mesa}. For consistency, the synthetic MR distribution at young ages ($<$100\,Myr) shown in this work is that obtained employing the higher entropy value.
\subsection{Stellar evolutionary models}\label{sec::model::mors}
The planetary atmospheric evolution models described in Section~\ref{sec::model::model} require as input also the stellar properties as a function of time to estimate the evolution of the atmospheric mass-loss rates and equilibrium temperature. In particular, the input should include the stellar bolometric luminosity ($L_{\rm bol}$) and the XUV luminosity ($L_{\rm XUV}$).%
Our framework employs the Mors stellar evolution code \citep{johnstone2021mors}, which predicts the evolution of the stellar rotation period and translates it into $L_{\rm XUV}$. In terms of stellar physics, the Mors code relies on the isochrones published by \citet{spada2013}.

As stars can be born with very different initial rotation rates \citep[e.g.][]{tu2015}, at early ages both the rotation period, and thus $L_{\rm XUV}$, can spread over more than an order of magnitude. In addition to comparing the relative input from initial atmospheric mass fractions and consequent atmospheric loss, one of the aims of this work is studying to which extent this spread can affect the evolution of planetary atmospheres across a wide parameter range and whether this spread has a noticeable impact in the observed MR distribution. Therefore, to model the evolution of the planetary atmospheres we employ three different evolutionary scenarios of the host star, namely that it evolved as a slow, a medium, or a fast rotator. As in \citet{kubyshkina2021mesa}, to account for the full possible spread, we consider for a solar mass star the evolutionary tracks corresponding to rotation periods of 1, 5, and 15\,days at an age of 150\,Myr.

Throughout, we consider just one possible host star of 1\,$M_{\odot}$, though the observed systems comprise also planets orbiting stars more and less massive than 1\,$M_{\odot}$. We limit this work to a 1\,$M_{\odot}$ host star, because the number of known exoplanets with both mass and radius measured with sufficient accuracy is not large enough to enable splitting the sample into smaller subsets to account for different host star masses and yet yield a meaningful comparison with model predictions. However, planetary atmospheric evolution depends to some extent on the mass of the host star. For example, \citet{kubyshkina2021mesa} demonstrated that at the orbital separations corresponding to the same equilibrium temperatures around stars of different masses, planets orbiting 1\,$M_{\odot}$ stars preserve more of their primordial atmospheres in comparison to planets evolving around lower-mass stars. Also, due to the specifics of stellar evolution, the difference between fast and slowly rotating stars is more pronounced for heavier stars \citep[e.g.][]{johnstone2021mors}. Therefore, accounting for observational uncertainties, the comparison we present here remains adequate for planets orbiting stars with masses lying in the 0.8--1.2\,$M_{\odot}$ range (that of the majority of the observed systems considered in this work; Figure~\ref{fig::obs_hist}), while for planets orbiting lower mass stars one should expect slight deviations from our model predictions towards smaller atmospheres (i.e. planetary radii) and less spread in the MR distribution as a result of the rotation history of the host star.

\section{The observed mass-radius distribution of intermediate-mass planets}\label{sec::observations}
\begin{figure}
   \centering
   \includegraphics[width=\hsize]{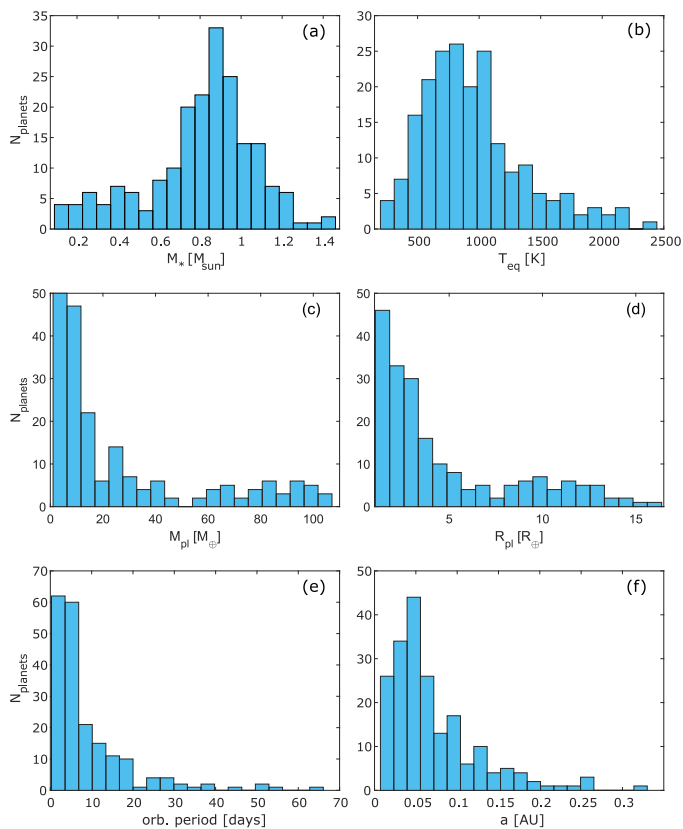} 
   \caption{Distribution of host star mass (a), equilibrium temperature (b), planetary mass (c), planetary radius (d), orbital period (e), and semi-major axis (f) for the considered data set of observed systems.}
   \label{fig::obs_hist}%
\end{figure}
For the comparison with the modeling results, we extract the sample of detected and confirmed planets from the NASA Exoplanet Archive\footnote{https://exoplanetarchive.ipac.caltech.edu/} employing the following filters. Since we do not restrict ourselves to Sun-like stars, we considered a range of orbital separations somewhat wider than that of the theoretical models by including planets with semi-major axes up to 0.4\,AU. We further considered planets with masses in the 1--110\,\Mer\ range and required the planetary radius to be measured. Finally, we filtered the selected planets by considering just those with relative uncertainties on the planetary mass and radius smaller than 45\% and 15\%, respectively. The resulting data set comprises 198 planets. Figure~\ref{fig::obs_hist} shows the data set distributions in terms of host star mass, equilibrium temperature, planetary mass, planetary radius, orbital period, and orbital semi-major axis.

\citet{kubyshkina2021mesa} showed that when comparing planets orbiting stars with different masses, one should rely on the equilibrium temperature rather than on the orbital separation when comparing with models. To ensure consistency, we do not consider the equilibrium temperature indicated in the database, but compute it as 
\begin{equation}
    T_{\rm eq} = T_{\rm eff}(1-\alpha)^{0.25}\left(\frac{R_*}{2a}\right)^{0.5}\,,
\end{equation}
where $\alpha$ is the planetary albedo, which we consider being equal to zero, $a$ is the orbital separation, and $T_{\rm eff}$ and $R_*$ are respectively the stellar effective temperature and radius extracted from the database. In case $T_{\rm eff}$ and/or $R_*$ were not available in the database, we extracted the missing information from \citet{TESS2019}.

Figure~\ref{fig::obs_hist} indicates that most of the host stars in the considered data set have masses between 0.6 and 1.2\,$M_{\odot}$, with the peak lying around 0.9\,$M_{\odot}$ and that most of the planets have equilibrium temperatures ranging between 500 and 1500\,K, which is comparable with that of the set of modelled planets, where $T_{\rm eq}$ ranges from 512\,K ($a$\,=\,0.35\,AU) to 1700\,K ($a$\,=\,0.03\,AU). In terms of mass and radius, the set of considered planets is dominated by planets with masses below 20\,\Mer\ and radii below 5\,\Rer, which are expected to be significantly affected by atmospheric mass loss. Above these values, the mass distribution looks nearly uniform (except for a noticeable lack of planets with masses between 40 and 50\,\Mer), while the radius distribution presents a secondary peak around 10\,\Rer\ (i.e. about Jupiter's radius), which is mainly due to planets more massive that $\sim$60\,\Mer\ and $T_{\rm eq}$ around 1000\,K. Driven by selection biases, most planets in the sample have short orbital periods, and thus short orbital separations.

For some of the planets (e.g. WASP-107\,b, GJ\,436\,b), the NASA exoplanet archive provides more than one entry. In these cases, to finally obtain a single entry for each parameter, we computed the median value taking the largest error bar given by the single entries as uncertainty.
\section{Comparing model predictions with observations}\label{sec::compare}
\subsection{Overall comparison}\label{sec::compare::overview}
Before proceeding with the comparison, for clarity we summarise here the properties of the database of theoretical models. It consists of a set of individual atmospheric evolutionary tracks for planets starting their evolution with 20 different masses, at five different orbital separations, and around three 1\,$M_{\odot}$ stars each with a different rotation history. In addition to these parameters, we consider two scenarios for the initial atmospheric mass fractions, where $f_{\rm at,0}$ increases with increasing planetary mass: the ``basic setup'' scenario with more compact initial atmospheres (in total 300 evolutionary tracks) and the ``migrated'' scenario, which assumes more voluminous initial atmospheres (in total 100 evolutionary tracks assuming the host star evolves as a slow rotator, because we employ this scenario just to outline the upper boundary of the MR distribution).

All evolutionary tracks have been run up to 10\,Gyr, and thus by combining all of them, we can construct the theoretical MR distribution at any given moment in time between protoplanetary disk dispersal (5\,Myr) and 10\,Gyr. To compare the modelling results with the observations, we consider an age of 5\,Gyr, which is about the average age of the observed planets. Although for most systems the age is poorly constrained, the MR distribution does not vary significantly at ages older than $\sim$3\,Gyr, and thus the particular choice of the comparison age does not modify the conclusions.

   \begin{figure*}[h]
   \centering
   \includegraphics[width=0.8\hsize]{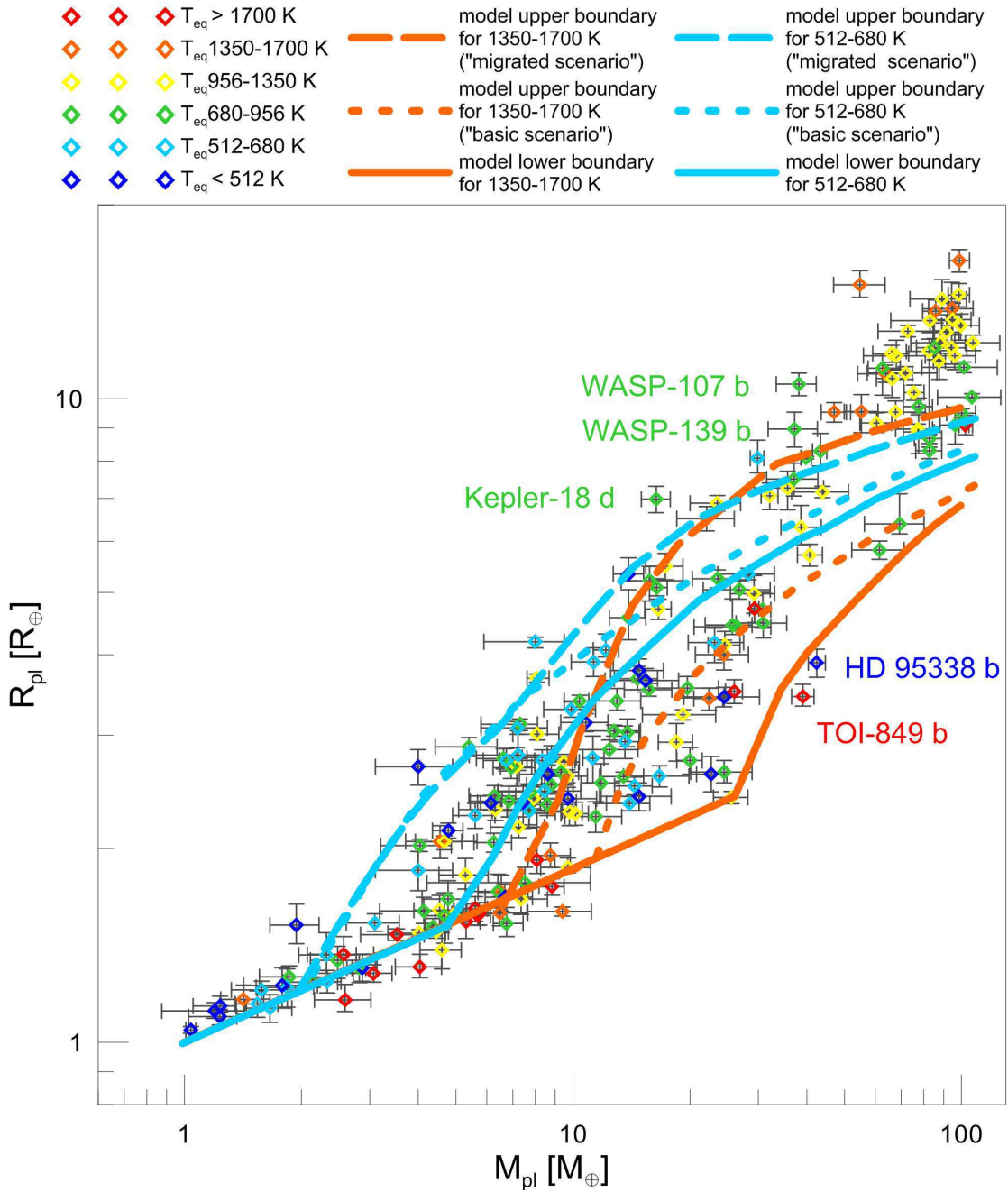} 
   \caption{MR distribution of the selected observed planets in comparison to modeling results. The symbol color indicates the approximate planetary equilibrium temperature as given in the legend. The largest uncertainties are 45\% on planetary mass and 15\% on planetary radius. \textcolor{dk}{The light blue and orange solid and long-dashed lines enclose the maximum predicted spread in planetary radius at the age of 5\,Gyr obtained considering the entire set of atmospheric evolutionary tracks computed in the 512--680\,K and 1350--1700\,K temperature intervals, respectively, where the upper boundary is given by the ``migrated'' scenario. For reference, the short-dashed lines show the upper boundary predicted for the ``basic setup'' scenario.}}
   \label{fig::MR_teq}
   \end{figure*}
\begin{table*}[h]
    \centering
    \textcolor{dk}{
    \caption{Model boundaries for each considered $T_{\rm eq}$ interval. The last column lists the number of observed planets falling within each $T_{\rm eq}$ interval.}
    \label{tab::boundaries}
    \begin{tabular}{c|c|c|c}
         \hline 
         \hline 
         $T_{\rm eq}$  & lower boundary & upper boundary & $N_{\rm points}$ \\
         \hline 
         $< 512$\,K    & 512\,K, fast rotator & -- & 22 \\
         512--680\,K   & 680\,K, fast rotator & 512\,K, slow rotator & 28 \\
         680--956\,K   & 956\,K, fast rotator & 680\,K, slow rotator & 63 \\
         956--1350\,K   & 1350\,K, fast rotator & 956\,K, slow rotator & 54 \\
         1350--1700\,K   & 1700\,K, fast rotator & 1350\,K, slow rotator & 16 \\
         $>1700$\,K   & -- & 1700\,K, slow rotator & 15 \\
    \end{tabular}
    } 
\end{table*}

Figure~\ref{fig::MR_teq} shows the comparison between the observed and synthetic MR distributions. \textcolor{dk}{We divide the 199 planets of the observed population into six $T_{\rm eq}$ intervals listed in Table~\ref{tab::boundaries}. Therefore, for planets with $512\,{\rm K} < T_{\rm eq} < 680\,{\rm K}$, we consider as lower theoretical radius boundary (solid light blue line in Figure\,\ref{fig::MR_teq}) the smallest $R_{\rm pl}$ as a function of planetary mass obtained in our models at temperatures of 512\,K (0.35\,AU) and 680\,K (0.2\,AU). This lower boundary is given by synthetic planets with $T_{\rm eq}$\,=\,680\,K orbiting the fast rotating star, because their initial atmospheres are smaller at 680\,K than at 512\,K (see Figure\,\ref{fig::initials}) and mass loss is greater for closer-in planets and in the case of a fast rotator. Similarly, the upper radius boundary in this temperature range is set by the planets with lower $T_{\rm eq}$ (i.e. 512\,K) orbiting the slowly rotating star. We also consider two upper boundaries, the one predicted by the ``basic setup'' scenario (short-dashed blue line in Figure\,\ref{fig::MR_teq}) and the one predicted by the ``migrated'' scenario (long-dashed light blue line in Figure\,\ref{fig::MR_teq}), where we consider the latter giving the predicted absolute maximum radius. We then apply the same scheme for all other temperature intervals (Table~\ref{tab::boundaries}).
The observed planets with $T_{\rm eq} < 512$\,K and $T_{\rm eq} > 1700$\,K lie outside of the boundaries of our synthetic population and for them we only give the lower and the upper radius limits, respectively.}

\textcolor{dk}{For each temperature interval, the lines drawn in Figure~\ref{fig::MR_teq} give also the approximate location of the regions of the MR distribution controlled by atmospheric mass-loss or by atmospheric accretion or by a mix of the two. Where the short-dashed (upper radius boundary within the ``basic setup'' scenario) and long-dashed lines (upper radius boundary within the ``migrated'' scenario) overlap, meaning that the upper radius boundary is roughly independent of the initial atmospheric mass fraction, the evolution is mostly controlled by atmospheric mass loss. Instead, deviations between the short- and long-dashed lines indicate that atmospheric accretion plays a role in the evolution. Therefore, for each planetary mass, the distance of the short- to the long-dashed line with respect to the distance of the short-dashed to the solid line is proportional to the relative importance of atmospheric accretion over mass loss on the evolution.}

The spread in planetary radius obtained from the evolutionary tracks employing the ``basic setup'' scenario resembles well the observed MR distribution for planets up to 15--20\,\Mer. This corresponds to the sub-Neptune mass planet population, whose evolution is believed to be mainly shaped by atmospheric escape. For these planets, the initial atmospheric mass fractions are of less importance compared to what occurs for the more massive planets. This can be seen by comparing the light blue short-dashed and light blue long-dashed lines in Figure~\ref{fig::MR_teq}, which show the final position of planets starting with very different initial conditions (``basic setup'' vs ``migrated'' scenarios) and evolving at 0.35\,AU ($T_{\rm eq} = 512$\,K) around the slowly rotating star. These two lines converge at about 7\,\Mer, reflecting the fact that below this mass the differences of a factor of a few in the initial atmospheric mass fractions does not have an impact on the final distribution. This effect was considered in detail by \citet{kubyshkina2020mesa}. The convergence point between the two regimes shifts towards higher masses with decreasing orbital separation and reaches $\sim$12\,\Mer\ at 0.03\,AU ($T_{\rm eq} = 1700$\,K, see Figure\,\ref{fig::MR_teq_apx_1350-2500}). Remarkably, the latter value is smaller than the heaviest bare core predicted in the ``basic setup'' scenario ($\sim20$~\Mer). This means that at such a short orbital separation and at an age of a few Gyrs planets with a present-day mass of 10--20\,\Mer\ can host a hydrogen-dominated atmosphere only if they started their evolution with a massive envelope. For example, the borderline case of a planet with a current mass of 12\,\Mer\ started its evolution as a 21.7\,\Mer\ planet comprising a 40\% initial atmospheric mass fraction. After 5\,Gyr this planet has lost almost entirely its primary atmosphere and it is still experiencing significant mass loss, which would lead to complete atmospheric loss within an age of 6\,Gyr.

Therefore, for planets more massive than $\sim$10\,\Mer, the initial conditions appear to be important for explaining the observed radius spread in the MR distribution, which requires higher initial atmospheric mass fractions than those estimated in the ``basic setup'' scenario. Indeed, considering the ``migrated'' scenario (where $f_{\rm at,0}$ varies from about 3\% for a 1\,\Mer\ planet to 70\% for a 108.6\,\Mer\ planet) the theoretical prediction fits well most of the observed MR distribution, except notably for the group of highly inflated hot planets with masses above $\sim$60\,\Mer\ that we discuss in Section~\ref{sec::results::outliers}.

\subsection{Radius spread}\label{sec::compare::spread}
\begin{figure*}
   \centering
   \includegraphics[width=\hsize]{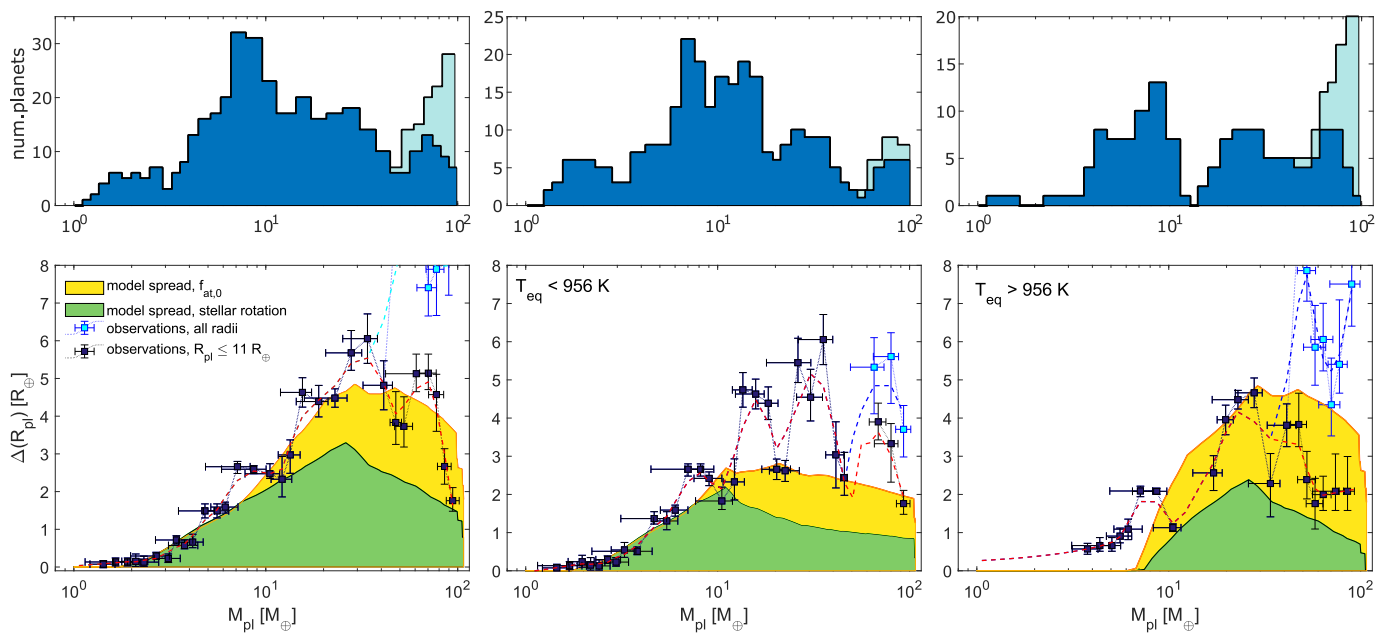} 
   \caption{Bottom row: comparison between the measured (squares) and synthetic (areas) spread in planetary radii as a function of planetary mass ($\Delta R_{\rm pl}$\,=\,$R_{\rm pl,max}-R_{\rm pl,min}$) for the entire sample (left), for planets cooler than 1000\,K (middle), and for planets hotter than 1000\,K (right). The relative values for the observed $\Delta R_{\rm pl}$ distribution are listed in Table~\ref{tab:spread_allpoints}, Table~\ref{tab:spread_allpoints_lowTeq}, and Table~\ref{tab:spread_allpoints_highTeq}, respectively, but with the difference that, for visualisation purposes, in the plots we combined neighbouring bins having $\Delta R_{\rm pl}$ values differing by less than 1\%. In the high-mass range, the blue and black squares have been obtained respectively accounting and disregarding planets affected by atmospheric inflation (i.e. planets with a radius larger than 11\,$R_{\oplus}$). The green area is for the ``basic'' scenario and it accounts for the variability in the initial stellar rotation rate affecting atmospheric mass loss. The yellow area is for the ``migrated'' scenario and thus it accounts for larger initial atmospheric mass fractions (assuming the host star was a slow rotator). To guide the eye, the red dashed line is a smooth of the distribution given by the black squares. Top panels: histograms of the number of observed planets within each mass bin for the entire sample (left), for planets cooler than 1000\,K (middle), and for planets hotter than 1000\,K (right). The dark blue and bright blue areas do respectively account and disregard planets affected by atmospheric inflation.}
   \label{fig::SPREAD_vs_mass}
   \end{figure*}

To aid understanding the results described above, we present in Figure~\ref{fig::SPREAD_vs_mass} the comparison between the \textcolor{dk}{observed $\Delta R_{\rm pl,obs}$ and synthetic $\Delta R_{\rm pl,model}$ radius spread} values as a function of planetary mass. In case of the observed MR distribution (black squares in Figure~\ref{fig::SPREAD_vs_mass}), we derived $\Delta R_{\rm pl,obs}$ as follows. We first divided the considered 1--110\,$M_{\oplus}$ mass range into 50 bins equally distributed in logarithmic space and computed the running median mass of the planets falling into every five consecutive bins. To each median mass value, we associated an uncertainty corresponding to the median error bar of the planets within each mass bin. For each mass bin, we then derived $\Delta R_{\rm pl,obs}$ by computing the difference between the radii of the largest ($R^{\rm obs}_{\rm pl,max}$) and smallest ($R^{\rm obs}_{\rm pl,min}$) planets. We then derived the uncertainty on $\Delta R_{\rm pl,obs}$ ($\sigma_{\Delta R_{\rm pl,obs}}$) for each mass bin by computing the difference between $\Delta R_{\rm pl,obs}$ and the difference between the radius of the largest planet, plus the 1$\sigma$ upper bound error bar ($\sigma_{R_{\rm pl,max}}^{\rm up}$), and the radius of the smallest planet, minus the 1$\sigma$ lower bound error bar ($\sigma_{R_{\rm pl,min}}^{\rm down}$), that is the sum of $\sigma_{R_{\rm pl,max}}^{\rm up}$ and $\sigma_{R_{\rm pl,min}}^{\rm down}$
\begin{eqnarray}
\sigma_{\Delta R_{\rm pl,obs}} &=& [(R^{\rm obs}_{\rm pl,max} +\,\sigma_{R_{\rm pl,max}}^{\rm up})\,-\,(R^{\rm obs}_{\rm pl,min} -\,\sigma_{R_{\rm pl,min}}^{\rm down})]\,-\,\Delta R_{\rm pl,obs}\nonumber \\
&=& \sigma_{R_{\rm pl,max}}^{\rm up} +\,\sigma_{R_{\rm pl,min}}^{\rm down}\,.
\end{eqnarray}
Table~\ref{tab:spread_allpoints} lists for each mass bin the $\Delta R_{\rm pl}$ values, and their uncertainties, obtained from the observed MR distribution, while Table~\ref{tab:spread_allpoints_lowTeq} and Table~\ref{tab:spread_allpoints_highTeq} give the $\Delta R_{\rm pl}$ values obtained considering planets with $T_{\rm eq}$ smaller and higher than 956\,K, respectively.

Within the synthetic MR relation, $\Delta R_{\rm pl,model}$ is controlled by planets lying at different orbital separations, thus having different $T_{\rm eq}$ values and receiving different amounts of stellar XUV flux throughout their evolution, implying that they experience different mass-loss rates. \textcolor{dk}{Therefore, at each planetary mass, $\Delta R_{\rm pl,model}$ is defined as the difference between the upper radius boundary at 512\,K and the lower radius boundary at 1700\,K (see Table~\ref{tab::boundaries}) as follows
\begin{equation}\label{eq::spread_synth}
    \Delta R_{\rm pl, model} = R^{\rm model}_{\rm pl,max}(512\,{\mathrm K})-R^{\rm model}_{\rm pl,min}(1700\,{\mathrm K})\,. 
\end{equation}
For any other temperature interval, we define $\Delta R_{\rm pl, model}$ following Equation~\ref{eq::spread_synth} substituting $R^{\rm model}_{\rm pl,max}(512\,{\mathrm K})$ and/or $R^{\rm model}_{\rm pl,min}(1700\,{\mathrm K})$ with those of the considered temperatures. In each interval, we consider $R^{\rm model}_{\rm pl,max}$ obtained employing both ``basic setup'' (green areas in bottom panels of Figure~\ref{fig::SPREAD_vs_mass}) and ``migrated'' scenario (yellow areas in bottom panels of Figure~\ref{fig::SPREAD_vs_mass}).} 

\textcolor{dk}{By default, the $\Delta R_{\rm pl, model}$ values include the impact on planetary atmospheres of the different evolution of the stellar rotation rate, and thus of the different amounts of stellar XUV radiation emitted during the early evolutionary stages.} The part of $\Delta R_{\rm pl,model}$ corresponding exclusively to the different possible evolution of the stellar rotation rate varies at different orbital separations and reaches its maximum value of $\sim1.1$\,\Rer\ at the innermost orbit of 0.03\,AU for planetary masses of $\sim26$\,\Mer. The maximum spread at other orbital separations, which is $\sim1.0$\,\Rer\ at 0.1\,AU and $\sim0.5$\,\Rer\ at 0.35\,AU, is achieved at smaller masses, namely at $\sim11$\,\Mer\ at 0.1\,AU and at $\sim3$\,\Mer\ at 0.35\,AU. In all cases, these are masses close to the heaviest bare cores predicted by our models, while the spread due to the differences in stellar histories decreases towards larger planetary masses due to the decreasing impact of escape. However, for the innermost orbit of 0.03\,AU this spread remains of $\sim0.5$\,\Rer\ around 80\,\Mer. Thus, the total $\Delta R_{\rm pl,model}$ within the ``basic setup'' scenario represents the sum of the spread due to the different stellar rotation histories with the spread caused by the different orbital separations, where the latter includes the effect of differences in irradiation levels and, where relevant (above $\sim10$\,\Mer), in initial atmospheric mass fractions. This total spread is shown by the green area in Figure~\ref{fig::SPREAD_vs_mass}.

Both Figure~\ref{fig::MR_teq} and \ref{fig::SPREAD_vs_mass} show that $\Delta R_{\rm pl}$ is not uniform as a function of planetary mass. For the lowest mass planets that have completely lost their atmospheres within 1\,Gyr $\Delta R_{\rm pl}$\,$\approx$\,0, it is largest for the intermediate-mass planets that keep some of their atmosphere in spite of experiencing significant atmospheric escape, and moderate for the heavier planets that are only weakly affected by atmospheric escape (if one excludes planets with radii larger than 11\,$R_{\oplus}$ that are affected by atmospheric inflation; see Section~\ref{sec::results::outliers}). In terms of atmospheric escape mechanisms, the atmosphereless planets typically lose their atmospheres already within the first hundreds of Myrs \citep[e.g.][]{kubyshkina2018grid}. For the intermediate-mass and massive planets, escape is mostly driven by the incident stellar XUV radiation, though for the latter group atmospheric evolution is mostly driven by thermal contraction, rather than atmospheric escape.

Figure~\ref{fig::SPREAD_vs_mass} shows that the $\Delta R_{\rm pl,model}$ predicted by the ``basic setup'' scenario is not a good fit to the observed radius spread (at least, for more massive planets), which is why we further considered the impact of \textcolor{dk}{possible planet migration, through considering the ``migrated'' scenario (Section~\ref{sec::model::initials}),} on $\Delta R_{\rm pl}$. 
As a matter of fact, within the ``migrated'' scenario the upper boundary of the radius spread is pushed upwards (yellow area in Figure~\ref{fig::SPREAD_vs_mass}) leading to a significantly better fit with the observations, particularly when considering the entire sample and excluding the planets subject to atmospheric inflation. However, the quality of the fit decreases somewhat by splitting the sample \textcolor{dk}{between low- ($\leq$956\,K) and high-temperature ($>$956\,K) planets.} 

In the low-temperature range, the synthetic $\Delta R_{\rm pl}$ distribution is a good fit to the observations up to about 12\,$M_{\oplus}$, while it is underestimated at higher masses. \textcolor{dk}{This discrepancy is caused by a group of dense Neptune-mass planets (comprising about 12\% of the observation data set) that lie outside of the model prediction, while the the vast majority agrees with the models;} we discuss thoroughly these outliers in Section~\ref{sec::results::outliers}.

In the high-temperature range, instead, the models underestimate slightly the observed $\Delta R_{\rm pl}$ distribution up to about 9\,$M_{\oplus}$ and slightly overestimate it at higher masses. Additional runs indicated that the reason of the mismatch at low mass \textcolor{dk}{can be} due to the fact that we considered a single stellar mass for the host star, while several planets in this temperature and mass range obit stars significantly more or less massive than 1\,$M_{\odot}$. This topic has been studied in detail by \citet{kubyshkina2021mesa}. \textcolor{dk}{Furthermore, for the lower mass planets our predictions are sensitive to some of the model assumptions, such as core density, which alone can account for a radius spread of about 0.5\,\Rer\ (see e.g. Figure~\ref{fig::MR_teq_apx_1350-2500}), and age (see e.g. Section~\ref{sec::results::outliers}).}
Instead, the one should not consider the mismatch at higher masses to be a concern for the following reasons. First, the ``migrated'' scenario sets the absolute upper limit and is not intended to exactly fit the observed distribution. \textcolor{dk}{Second, the majority of planets with $T_{\rm eq}>956$\,K are concentrated near or above the upper boundaries predicted for the ``basic setup'' scenario, suggesting that the majority of planets in this region would follow the ``migrated'' scenario. Given that this lack of planets within the area predicted by the ``basic setup'' scenario pushes upwards the lower boundary of the observed population, one should expect to see points within the yellow area.} Finally, for planets more massive than 50\,\Mer\, the position of the observed distribution depends on where one sets the boundary at which atmospheric inflation set in, which we set rather arbitrarily. 
\subsection{Outliers}\label{sec::results::outliers}
\textcolor{dk}{In total, we find that about 20\% of the 199 observed planets we considered lie outside of the boundaries drawn by the synthetic models. About half of them are hot and massive planets with radii above 10\,\Rer\ (top-right corner of Figure\,\ref{fig::MR_teq}). Besides for a few exceptions, the remaining outliers are dense and relatively cool Neptune-mass planets, whose fraction increases with decreasing temperature: we find only 3 of them in 680--956\,K interval (of 63 planets in total), but the number increases to 6 out of 28 planets in the 512--680\,K interval and to 11 out of 22 for $T_{\rm eq} < 512$\,K.}

As shown by Figure~\ref{fig::MR_teq} and \ref{fig::SPREAD_vs_mass}, the group of planets most diverging from the model predictions are sub-Saturns, with masses between 60 and 110\,\Mer\ and temperatures in the 800--1500\,K range. For these planets, the radii exceed 10--11\,\Rer, reaching 12\,\Rer\ for $T_{\rm eq}<1000$\,K and 16.4\,\Rer\ for $T_{\rm eq}$ between 1350 and 1500\,K. Furthermore, the radii of these planets show a clear correlation with equilibrium temperature and planetary mass. Even within the ``migrated'' scenario and for 60--110\,\Mer\ planets (these planets present 60--70\% of their mass in the atmosphere), such large radii are achieved only within the first 10\,Myr of evolution, while planets are hot, but none of the observed planets in this group is younger than 1\,Gyr. %
Although for these planets atmospheric mass loss is negligible (it reaches up to $\sim10^{12}$\,g\,s$^{-1}$ in the first Gyr of evolution, but it is unable to remove even 1\% of the initial atmosphere), thermal contraction alone pushes the planetary radii below 10\,\Rer\ within a few tens of Myr, independently on the assumed initial entropy (i.e. core temperature). We even attempted to reproduce the observed distribution by increasing the initial atmospheric mass fraction to 90\%, but the resulting planetary radii hardly exceeded 10\,\Rer.%
\begin{figure}
   \centering
   \includegraphics[width=\hsize]{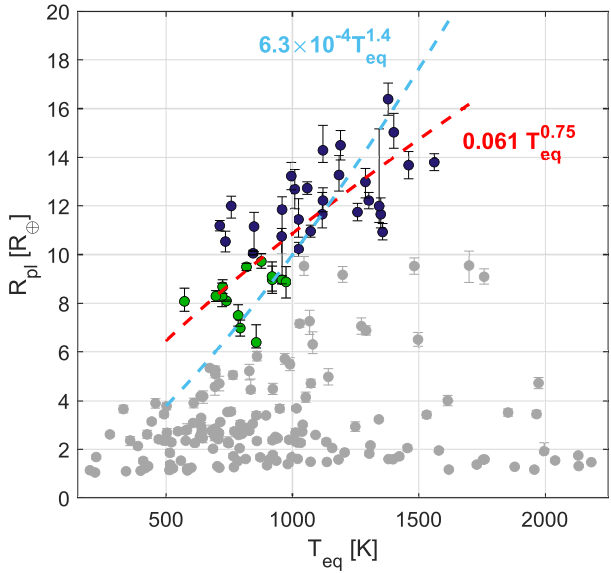} 
   \caption{Radii of the observed planets as a function of equilibrium temperature. The black points mark the group of ``inflated Saturns'' discussed in the text ($R_{\rm pl}\geq 10$\,\Rer), while the green points are planets with 6\,$\leq$\,$R_{\rm pl}$\,$<$\,10\,\Rer\ and $T_{\rm eq}<1000$\,K. All other planets are marked in gray. The red dashed line shows the fit to the black points, while the blue dashed line shows the fit to these planets assuming the exponent obtained for the inflated hot Jupiters \citep{Laughlin2011}.}
   \label{fig::R-on-Teq}
\end{figure}

This suggests that some additional heating mechanism (or cooling suppression mechanism) is at play in inflating the radii of these planets. Therefore, these objects might represent the low-mass edge of the population of inflated hot Jupiters \citep{Bodenheimer2001,guillot2002,baraffe2003,Laughlin2011,demory2011,Fortney2021HJ}. However, Figure~\ref{fig::R-on-Teq} shows that for planets in this sample ($R_{pl}>10$\,\Rer\ and $T_{\rm eq}>800$\,K) the dependence of radius on equilibrium temperature is about $R_{\rm pl}\sim T_{\rm eq}^{0.75}$, which is shallower than that found for hot Jupiters at temperatures above 1200\,K, that is $R_{\rm pl}\sim T_{\rm eq}^{1.4}$ \citep{Laughlin2011}. We remark that a more detailed analysis of the data indicates that the value of the exponent seems to depend on equilibrium temperature as well, but the still small size of the sample does not allow one to make any robust conclusion. Furthermore, the group of inflated planets might extend down to $\sim$500\,K if one includes also planets with radii of 6-10\,\Rer\ and equilibrium temperatures below 1000\,K (green points in Figure~\ref{fig::R-on-Teq}). With the inclusion of these smaller, lower-temperature planets, the lowest planetary mass in the group of inflated sub-Saturns decreases from 38.1\,\Mer\ to 16.4\,\Mer. 

For context, a wide range of possible mechanisms has been put forward to explain the radius inflation of hot Jupiters. Due to the strong correlation with $T_{\rm eq}$ (and stellar $T_{\rm eff}$), most theories suggest that atmospheric inflation is connected with star-planet interaction, which includes a number of specific processes that can be separated into three groups: tidal interactions with host star \citep[e.g.][]{Bodenheimer2001,Arras2010}, hydrodynamic heat transport towards planetary interior \citep[e.g.][]{Showman2002,Youdin2010,Tremblin2017,Sainsbury-Martinez2019}, and heating due to induced currents deep in the atmosphere \citep[Ohmic dissipation or induction heating;][]{Batygin2010,perna2010,wu2013,Ginzburg2016,kislyakova2018indheat,Knierim2022}.

\begin{figure*}[h]
   \centering
   \includegraphics[width=0.48\hsize]{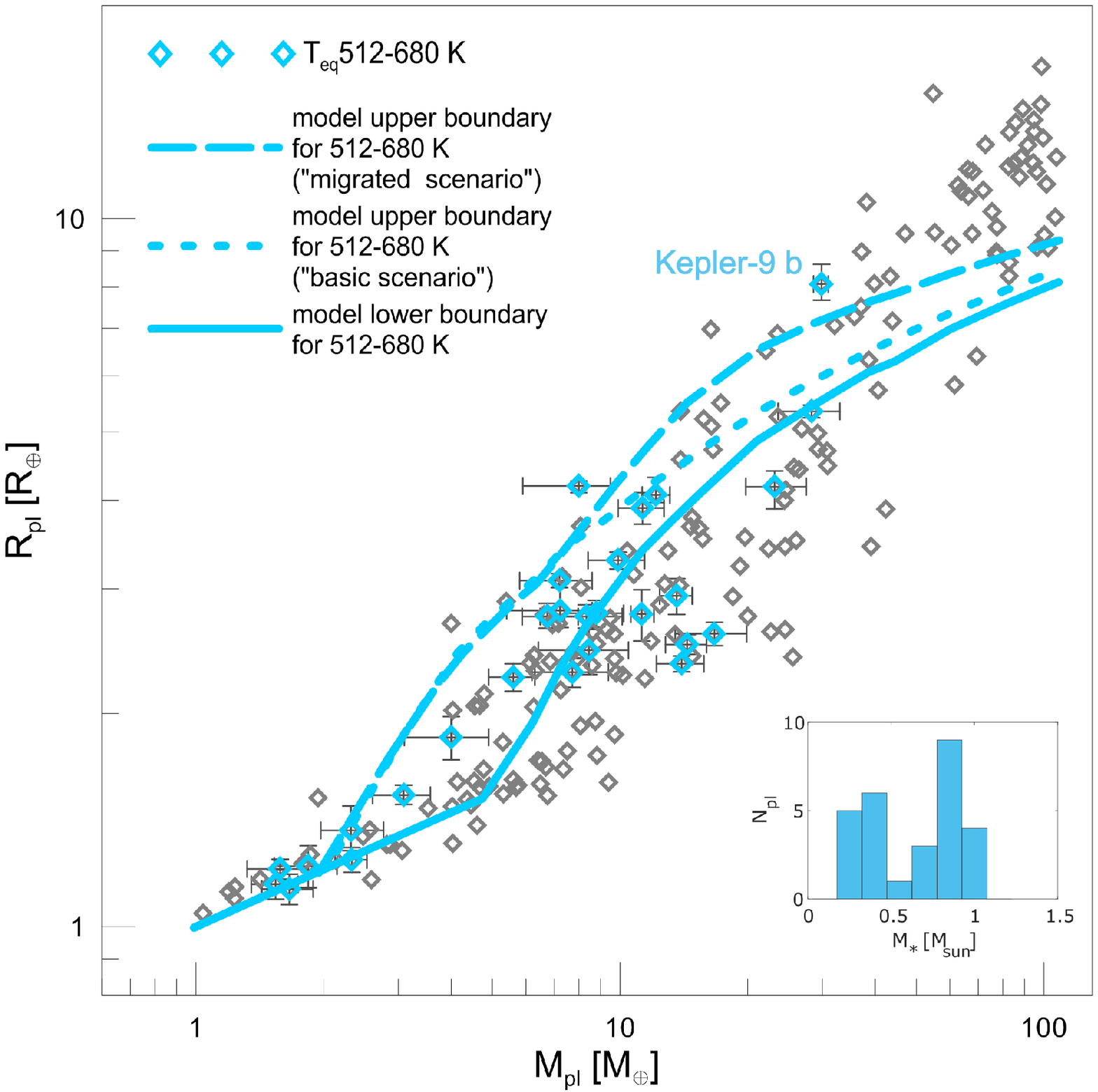} \includegraphics[width=0.48\hsize]{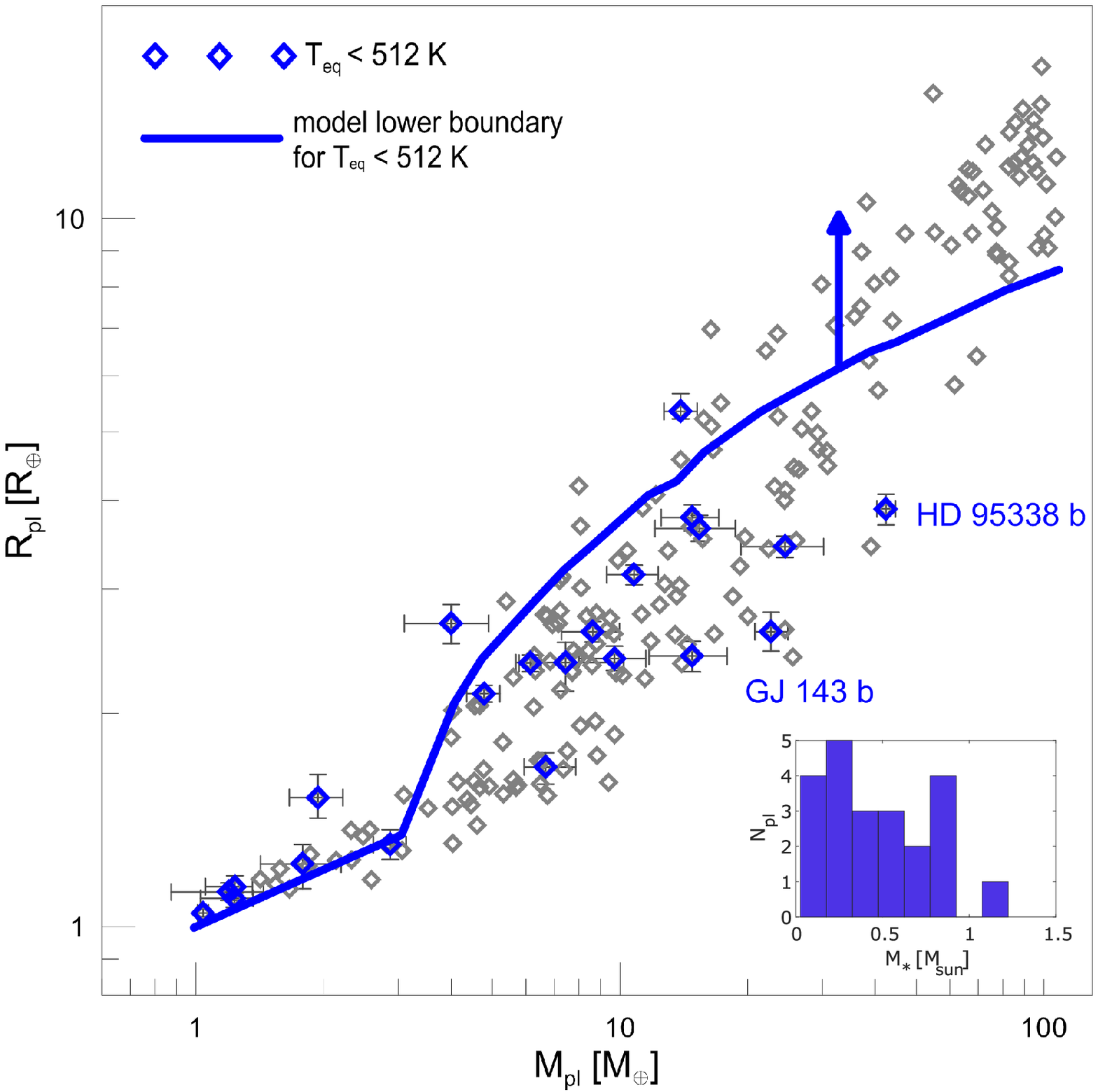} 
   \caption{\textcolor{dk}{Same as Figure~\ref{fig::MR_teq}, but for the 512--680\,K (left panel, light blue) and $<$512\,K (right panel, dark blue) temperature range. Planets from other temperature intervals, shown in black, are included in the plots for clarity. The light blue (left panel) and dark blue (right panel) solid lines indicate the lower boundary of the theoretical radius spread for the respective temperature intervals. The long-dashed and short-dashed light blue lines (left panel) indicate the upper boundaries of the radius spread in the 512--680\,K interval according to the ``migrated'' and ``basic setup'' scenarios, respectively. For context, the inset located in the bottom-right corner shows the mass distribution of the stars hosting the observed planets lying within each considered temperature range.} %
   }
   \label{fig::MR_teq_cool}%
   \end{figure*}

Besides the group of hot sub-Saturns, there are a few specific planets that diverge from the modeling predictions and stand out compared to the other observed planets with comparable mass and temperature. Figure~\ref{fig::MR_teq} highlights the most evident outliers that are WASP-107\,b, WASP-139\,b, Kepler-18\,d, TOI-849\,b, and HD\,95338\,b. The first three lie in the 680--956\,K equilibrium temperature bin and have a particularly low density, while the latter two are located below the lower radius boundary for planets at an orbit equivalent to $T_{\rm eq} = 1700$\,K.
\textcolor{dk}{Of these planets, WASP-139\,b is the youngest, with an estimated age of $\sim0.5$\,Gyr \citep{Hellier2017}; the parameters of this planets are in agreement with our predictions at the age of the system, and thus WASP-139\,b should not be considered to be an outlier. The same applies to the $2_{-1.3}^{+2}$\,Gyr-old planet Kepler-9\,c \citep[see Figure~\ref{fig::MR_teq_cool};][]{Borsato2019} and possibly to $\pi$\,Men\,c as well \citep[see Figure~\ref{fig::MR_teq_apx_956-1350};][]{Huang2018,Gandolfi2018}.}

\textcolor{dk}{TOI-849\,b is consistent with the theoretical prediction given by the upper radius boundary shown in Figure~\ref{fig::MR_teq_apx_1350-2500}. However, we highlight this planet, because it is considerably more dense than planets of similar mass and temperature. Figure~\ref{fig::MR_teq} and Figures~\ref{fig::MR_teq_apx_1350-2500}--\ref{fig::MR_teq_apx_956-1350} show that the majority of planets hotter than 1000\,K lie near, or somewhat above, the upper limit of the predicted radius spread within the ``basic setup'' scenario (see the short-dashed red, orange, and yellow lines).} This suggests that the relatively small initial atmospheres assumed in this scenario for hot planets are unlikely. To reproduce the MR distribution of planets hotter than 1500\,K, the initial atmospheres should necessarily be substantial, as assumed in the ``migrated'' scenario (see the long-dashed red and orange lines in Figure~\ref{fig::MR_teq_apx_1350-2500}), implying that these planets were formed further away from the star and migrated inwards, in agreement with recent formation theories \citep{morbidelly2016,jin2018,morbidelly2020}.
Instead, HD\,95338\,b has an equilibrium temperature of about 385\,K \citep{diaz2020hd95338}, and lies, therefore, far below the lower edge \textcolor{dk}{predicted for planets cooler than 512\,.} Figure~\ref{fig::MR_teq_cool} shows that this is not an isolated case as there is the tendency for planets on the ``cold'' side ($T_{\rm eq}<680$\,K) of the distribution to behave in a similar way.

\begin{figure}
    \centering
    \includegraphics[width=\hsize]{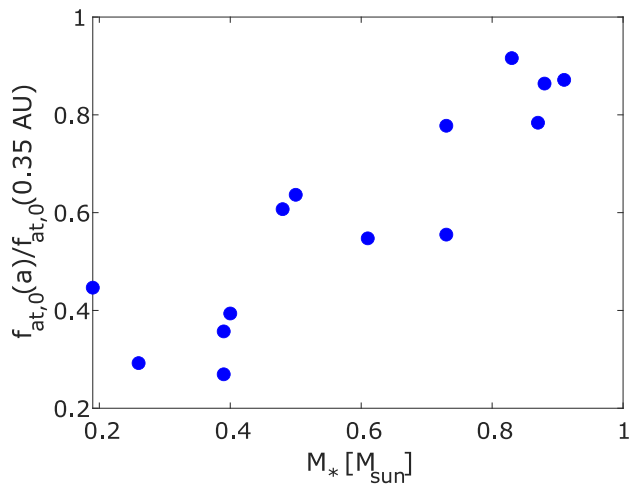}
    \caption{\textcolor{dk}{Ratio between $f_{\rm at,0}(a)$ and $f_{\rm at,0}(0.35 {\rm AU})$ as a function of host star mass for the observed planets with $T_{\rm eq}<512$\,K and $M_{\rm pl}$\,$>$\,6\,$M_{\oplus}$.}}
    \label{fig::fat_ms_trend_cool}
\end{figure}

Indeed, a significant fraction of the cooler planets ($<$680\,K) lies below the lower boundaries predicted by the model (Figure~\ref{fig::MR_teq_cool}). Planets at these temperatures are less affected by atmospheric escape than the hotter ones and their final position in the MR diagram is strongly dependent on their initial parameters. Therefore, one would expect to see such planets spreading around the theoretical predictions, because the initial mass fractions we employed are average values of those predicted by formations models. \textcolor{dk}{However, part of the outliers (especially for $T_{\rm eq}<512$\,K)} stand out significantly and their bulk density suggests that they can possess only small hydrogen-dominated atmospheres of less than 1--2\%, which is an order of magnitude lower than the typical atmospheric mass fractions expected for planets in this part of the parameter space (e.g. HIP\,97166\,b, K2-292\,b, K2-110\,b, GJ\,143\,b), though one cannot exclude the possibility of the model underpredicting mass-loss rates for these planets. 

\textcolor{dk}{There is also the possibility that these planets do not match our prediction as a result of our assumption that all planets orbit a 1\,$M_{\odot}$ star. As a matter of fact, the observational data set is dominated by planets orbiting stars with masses below 0.7\,$M_{\odot}$ for planets cooler than 680\,K and with masses below 0.5\,$M_{\odot}$ for planets cooler than 512\,K, as indicated by the histograms in Figure\,\ref{fig::MR_teq_cool}. Furthermore, the fraction of low-mass stars in the observational data set increases progressively with decreasing $T_{\rm eq}$ (see also Figures\,\ref{fig::MR_teq_apx_1350-2500}--\ref{fig::MR_teq_apx_680-956}) and the fraction of dense Neptunes lying below the boundaries predicted by our models behaves in the same way. The same temperatures are achieved at closer orbital separations for lower-mass stars implying that, at the same equilibrium temperature, planets orbiting lower mass stars will lose more atmosphere compared to those orbiting higher mass stars \citep[e.g.][]{kubyshkina2021mesa}. Also, we overestimated the results one would get applying Equation~(\ref{eq::fat_mordasini2020}) considering the orbital separations obtained taking into account the specific stellar masses. To illustrate this, we used Equation~(\ref{eq::fat_mordasini2020}) to compute $f_{\rm at,0}(a)$ that is the initial atmospheric mass fraction obtained using the actual orbital separations ($a$) for the observed planets with $T_{\rm eq}<512$\,K and $M_{\rm pl}$\,$>$\,6\,$M_{\oplus}$. We then used again Equation~(\ref{eq::fat_mordasini2020}) to compute $f_{\rm at,0}(0.35 {\rm AU})$ that is the initial atmospheric mass fraction obtained for the same planets, but assuming an orbital separation of 0.35\,AU (i.e. ``basic setup'' scenario). Finally, we plot in Figure~\ref{fig::fat_ms_trend_cool} the ratio between $f_{\rm at,0}(a)$ and $f_{\rm at,0}(0.35 {\rm AU})$ as a function of the mass of the host stars. The ratio decreases with decreasing stellar mass implying that to adequately model these planets one would need to employ an atmospheric accretion model suitable for low-mass stars. However, there are still a few outliers falling in the dense Neptunes category that orbit Sun-like stars (i.e. HD\,95338\,b, Kepler-411\,b, K2-292\,b).}
Appendix~\ref{apx::outliers} discusses thoroughly the single most important outliers from the theoretical MR distribution.

\section{Escape-dominated and formation-dominated regions of the MR diagram}\label{sec::discussion}
The analysis presented in Section~\ref{sec::compare::overview} shows that combining average primordial planetary parameters predicted by formation models with atmospheric evolution leads to describe well the observed overall MR distribution (see Figures\,\ref{fig::MR_teq} and \ref{fig::SPREAD_vs_mass}). Furthermore, our analysis highlights regions of the parameter space characterised by qualitatively different planetary origin and/or evolutionary path (Figures\,\ref{fig::MR_teq} and \ref{fig::MR_teq_cool}). 

For planets with present-day mass below $\sim10$\,\Mer, most of the radius spread is shaped by differences in atmospheric mass-loss rates caused by different orbital separations and different evolutionary tracks of the stellar rotation rate (see Figure\,\ref{fig::SPREAD_vs_mass}). Therefore, for these planets atmospheric loss is the dominant mechanism driving atmospheric evolution. The primordial parameters characterising these planets play a minor role in shaping the final MR distribution, as a wide range of possible initial parameters can lead to the same outcome \citep[see][]{kubyshkina2018grid,kubyshkina2019a,kubyshkina2020mesa}.

Instead, for more massive planets atmospheric loss becomes progressively less relevant, and the atmospheric evolution of planets heavier than 60\,\Mer\ is mainly driven by thermal contraction. In this case, the initial planetary parameters, in particular the initial atmospheric mass fraction, play the leading role. Therefore, to explain the upper part of the observed radius spread for planets more massive than 10\,\Mer, one has to assume that these planets were formed with large hydrogen-dominated envelopes (Figure\,\ref{fig::MR_teq}). This appears to be particularly important for planets in close-in orbits, where the radius spread corresponding to different initial conditions is up to four times broader than the maximum spread due to the different possible stellar evolutionary paths.

Finally, for orbits corresponding to $T_{\rm eq}\lesssim$800\,K, the radius spread displayed by $\sim 7-30$\,\Mer\ planets is significantly broader than predicted by the models, with a significant number of planets presenting unexpectedly (for their temperature) high densities. This group of high-density planets comprises about 6\% of the whole data set suggesting that they represent a systematic deviation from the theoretical prediction, rather than individual outliers (see Figure~\ref{fig::MR_teq_cool}). 

The atmospheric evolutionary models indicate that planets with $T_{\rm eq}\lesssim$800\,K and masses around 20--30\,\Mer\ are weakly affected by atmospheric escape (e.g. HD\,95338\,b), while the less massive planets can lose a substantial part of their atmosphere (e.g. K2-292\,b, see details in the Appendix). However, for planets in the 7--30\,\Mer\ range mass loss is not sufficient to entirely remove the primary hydrogen-dominated atmosphere, which is expected to be in the 7--40\% range in the ``basic setup'' scenario and to reach up to $\sim$60\% in the ``migrated'' scenario. Therefore, to reach radii comparable to those of the lower edge of the observed distribution, these planets had to start their evolution with atmospheric mass fractions similar to the present-day values (i.e. to have formed in a gas-poor environment) or to have lost their primary envelope as a result of some catastrophic event. However, the latter possibility is unlikely particularly because of the smooth deviation of the observed high-density planets from the rest of the population in the MR distribution (left panel of Figure~\ref{fig::MR_teq_cool}). This suggests that the underlying planetary structures, and thus the formation scenarios that led to them, in this temperature and mass range can be much more complex than that of a rocky core surrounded by a hydrogen-dominated envelope as considered in our models.

\begin{figure}
   \centering
   \includegraphics[width=\hsize]{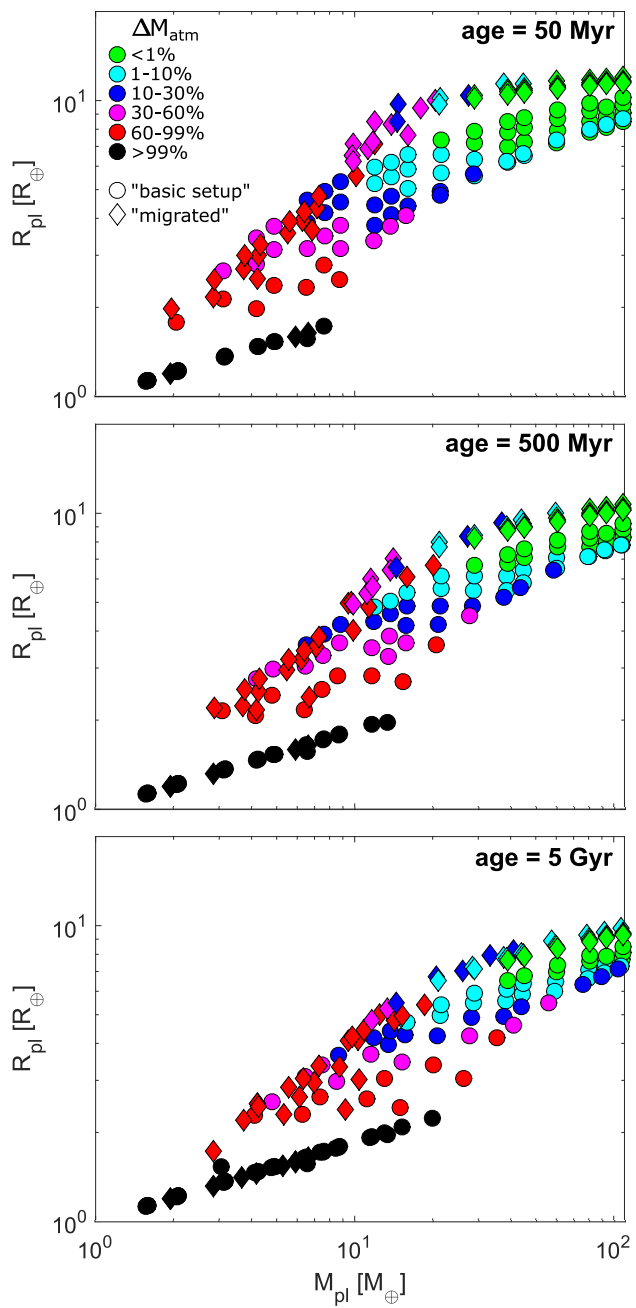} 
   \caption{Synthetic MR diagram color coded according to the fraction of primordial atmosphere that has been lost through escape at 50\,Myr (top), 500\,Myr (middle), and 5\,Gyr (bottom) across the entire model population (i.e. 0.03--0.35\,AU). The color code is defined in the legend located in the top-left corner of the top panel. Circles and diamonds correspond to the initial conditions set according to the ``basic setup'' and ``migrated'' scenarios, respectively.}
   \label{fig::MR_lost}%
\end{figure}

Therefore, the relative role of the specific formation mechanisms and of atmospheric evolution driven by thermal contraction and mass loss in shaping the MR distribution of intermediate-mass planets is not the same across the entire parameter space. Figure~\ref{fig::MR_lost} shows the impact of atmospheric escape (including both ``basic setup'' and ``migrated'' scenario) at 50\,Myr (near the end of the initial extreme mass-loss phase driven mainly by the own planetary thermal energy), 500\,Myr (when the XUV emission of a moderate rotator decreases by an order of magnitude), and 5\,Gyr in the different regions of the synthetic MR distribution (including all orbital separations considered in the model).

Already after 50\,Myr of evolution most planets with mass smaller than 10\,\Mer\ lose more than 30\% of their initial atmosphere, with a significant fraction of planets ending up with a bare core. This holds for planets within both scenarios (i.e. sets of initial parameters). Although the host star is most active at young ages, in the first 50\,Myr and within the ``basic setup'' scenario the atmospheric escape endured by planets less massive than 10\,\Mer\ is mainly driven by their own thermal energy and low gravity \citep[e.g.][]{kubyshkina2018grid,kubyshkina2021mesa}. In the ``migrated'' scenario, this is the case for planets less massive than $\sim$20\,\Mer (see the top panel of Figure\,\ref{fig::MR_lost}). Since a significant fraction of their mass is in the atmosphere ($\sim$25--40\% for the initial mass range of 9--20\,\Mer) and these planets are highly vulnerable to escape (as a result of their large radii), throughout their evolution these planets move significantly across the MR diagram towards lower masses and radii. %

Our models indicate that the initial phase of extreme escape is responsible for the formation of the radius gap \citep[e.g.][]{fulton2017,owen_wu2017} around 1.8--2\,\Rer\ already after 50\,Myr. Furthermore, Figure~\ref{fig::MR_lost} shows that the predicted position of the radius gap in the MR diagram is mass dependent and it moves towards larger radii with increasing mass. In particular, at an age of 50\,Myr the radius gap extends up to 8\,\Mer\ and it extends to higher masses with time (about 15\,\Mer\ after 500\,Myr and about 20\,\Mer\ at 5\,Gyr) under the action of atmospheric escape driven by the stellar XUV irradiation. This demonstrates that (1) most of the atmospheric escape occurs within the first tens to hundreds Myrs, but (2) the long-term XUV driven escape on Gyrs timescale plays a significant role in shaping the population of low- and intermediate-mass planets. At the end of the evolution, at 5\,Gyr, the planets that have been significantly affected by atmospheric escape (i.e. lost at least 30\% of their initial atmosphere) have masses as large as 60\,\Mer, though the majority have masses below 20\,\Mer. %

\begin{figure}
   \centering
   \includegraphics[width=\hsize]{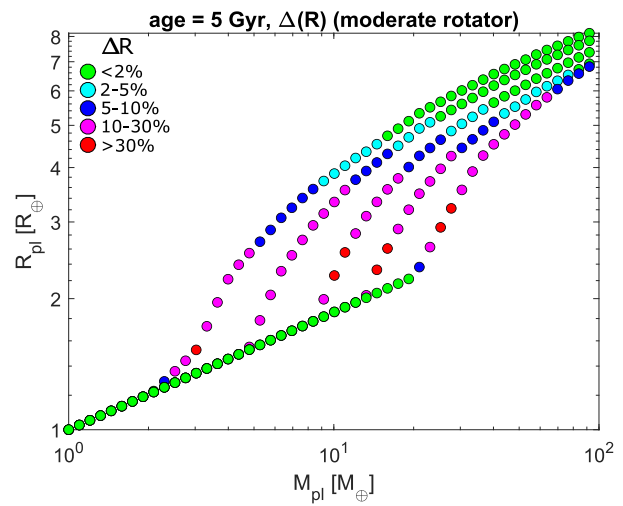} 
   \caption{Synthetic MR diagram color-coded according to the relative difference in planetary radii between planets that evolved around a slow, medium, and fast rotator at an age of 5\,Gyr \textcolor{dk}{(see Equation~\ref{eq::dR_rotations}), as indicated in the legend. The position of the points shown in Figure~\ref{fig::MR_DRrot} is that obtained for the ``basic setup'' scenario considering a moderate rotator following interpolation on a regular mass grid.}}
   \label{fig::MR_DRrot}%
    \end{figure}

Understanding, the relative role of atmospheric escape and formation mechanisms with the consequent thermal evolution in shaping the MR relation is also important for the analysis of the evolution of planet-hosting stars. \citet{kubyshkina2019a,kubyshkina2019kepler11} and \citet{bonfanti2021evol} developed an algorithm capable of constraining the past activity (rotation) history of stars hosting sub-Neptunes. The scheme is based on modeling the evolution of planetary atmospheres (similarly to the approach described in this study, but without thermal contraction) to constrain the planetary initial atmospheric mass fraction and evolution of the stellar rotation rate (a proxy for the stellar XUV emission) within a Markov-Chain Monte Carlo framework through fitting the present-day observed system parameters. One key prerequisite is that the radius of the planet should be sensitive to the activity history of the host star, which in turn requires that the planetary atmosphere is significantly affected by mass loss. However, there is no linear relation between differences in the lost atmospheric mass and differences in planetary radii between planets evolving around stars with different activity histories. The final differences in planetary radii depend additionally on mass and orbital separation of the planet \citep{kubyshkina2021mesa}. Figure~\ref{fig::MR_DRrot} shows how the relative difference in planetary radii caused by different rotation scenarios of the host star changes across the synthetic population within the ``basic setup'' scenario. Here,
\begin{equation}\label{eq::dR_rotations}
\Delta R = 100\%\,\frac{R_{\rm pl, slow} - R_{\rm pl, fast}}{R_{\rm pl, moderate}}\,,
\end{equation}
where $R_{\rm pl, slow/medium/fast}$ are the final (5\,Gyr) radii of the same planet orbiting around a slow/medium/fast rotator, respectively. The maximum $\Delta R$ in our synthetic population is of about 43\% and it exceeds 10\% for a significant part of the population. For it to enable constraining the history of the activity of the host star, $\Delta R$ should be larger than the measured uncertainty on the planetary radius. For comparison, 96\% of the detected planets considered in this work have a relative uncertainty on the planetary radius smaller than 30\%. 
%
\section{Conclusions}\label{sec::conclusions}
We have analysed the MR distribution of low- and intermediate-mass planets, comparing it to the predictions of theoretical models computed by joining MESA with the results of upper atmosphere hydrodynamic simulations to accurately estimate mass-loss rates. This grid of hydrodynamic models covers planets in the 1--110\,\Mer\ mass range and 300--2000\,K temperature range, which set up the applicability boundaries of our evolution models. To ensure that the comparison with the observed MR distribution is not significantly affected by observational uncertainties, we consider only planets with uncertainties smaller than 15\% in radius and 45\% in mass, which leaves us with a sample of 199 planets. Most of these planets orbit Sun-like stars and thus we considered all synthetic planets orbiting a 1\,$M_{\odot}$ star, but accounted for different stellar activity evolution scenarios. In terms of initial atmospheric mass fractions, we used predictions derived from detailed formation models as a function of planetary masses and orbital separations (``basic setup'' scenario) or assuming formation at 1.5\,AU and follow-up migration inside the disk (``migrated'' scenario).

Studying the population of detected exoplanets is key to constrain fundamental planetary formation and evolution processes (e.g. atmospheric accretion and escape, migration, core composition). So far, studies of the exoplanet population focused mostly on the planetary orbital period vs radius diagram, with the occasional addition of the stellar mass as a further parameter. Here, we demonstrated that, thanks to ongoing ground- and space-based surveys, the currently available exoplanet population enables one to consider also the planetary mass, which is a key parameter controlling several physical processes, such as atmospheric accretion, atmospheric escape, and migration. 

The observed MR distribution is shaped by a combination of planetary formation processes, which determine the properties of planetary systems during the first Myrs of evolution, and long-term atmospheric evolution driven by thermal contraction (slow dissipation of post-formation luminosity) and mass loss. Our evolution models lead to reproduce well the shape of the observed MR distribution, despite the use of coarse approximation for estimating the initial atmospheric mass fractions. Furthermore, we showed that the relative impact to the shape of the MR distribution given by the initial planetary parameters and the consequent evolution is not uniform across the observed population. The boundary between escape-dominated and formation-dominated regions of the MR distribution has a complex shape and cannot be attributed to one specific parameter, such as planetary mass or orbital separation. 

In general, the low-mass part ($M_{\rm pl}\leq 10$\,\Mer) of the population is fully dominated by atmospheric loss. All planets in this mass range are subject to extreme atmospheric escape driven by their own high thermal energy and low gravity throughout the first few tens of megayears of their evolution. This atmospheric loss becomes more intense with increasing planetary radius, and thus with increasing planetary core temperature and atmospheric mass fraction. Our simulations predict that increasing any of these two parameters at the time of the protoplanetary disk dispersal does not lead to significant changes in the final planetary properties after Gyrs of evolution \citep[see e.g.][for a detailed discussion]{kubyshkina2019a,kubyshkina2021mesa}. Therefore, recovering the initial parameters of these planets from their present-day properties is, in general, difficult.

For more massive planets ($M_{\rm pl}\geq10$\,\Mer), the initial atmospheric conditions and the specific formation processes that led to them become important. In particular, we had to assume that planets formed with substantial hydrogen-dominated atmospheres (potentially beyond the snow line) to explain the upper part of the MR distribution. In this part of the distribution, the maximum radius spread associated with atmospheric mass loss, including the impact of the different paths of stellar rotation evolution, is about one fourth of the entire spread. Furthermore, the relative contribution of the initial conditions on atmospheric evolution and thus on the radius spread depends on the orbital separation (i.e. equilibrium temperature). For planets cooler than $T_{\rm eq}\sim700$\,K, escape plays a significant role in atmospheric evolution up to 20--30\,\Mer, at $T_{\rm eq}\sim1000$\,K this mass boundary lies around 40--50\,\Mer, and for the hottest planets it moves up to 100\,\Mer. However, in all cases the impact of the initial conditions remains dominant, particularly for the hottest planets.

The atmospheric evolution models presented in this work explain the position in the MR diagram of about 80\% of the detected planets with well-measured masses and radii. The remaining 20\% is mostly distributed into two distinct groups of outliers. The first one consists of about 8.5\% of the total number of planets and it comprises inflated sub-Saturns in close orbits (2--6\,days) around their host stars, which require some kind of additional internal heat source to reproduce their radii that are larger than predicted by the model. The strong correlation between radius and equilibrium temperature displayed by these planets indicates that the additional heating mechanism should be similar, if not the same, as that inflating the radii of hot Jupiters. The detailed analysis of WASP-107\,b suggests that this additional heating mechanism can also be effective for relatively cool ($\sim700$\,K) and low-mass ($\sim30-40$\,\Mer) planets.

The second group of outliers consists of about 12\% of the total number of planets and it comprises dense sub-Neptune-size planets with $T_{\rm eq}\leq700$\,K and masses, in general, between 10 and 30\,\Mer. The simulations indicate that to achieve the observed planetary parameters at ages of a few Gyrs, \textcolor{dk}{the most extreme of these planets} had to be formed with atmospheric mass fractions comparable to those currently present, namely $\sim1-2$\%, which is an order of magnitude smaller than predicted by the atmospheric accretion models used in this study. \textcolor{dk}{At the same time, the mass-radius relation of these planets resembles that of icy cores \citep{zeng2021gap_waterEOS}, which could suggest that their internal structure is different from that of a rocky core surrounded by a hydrogen-dominated envelope, thus providing an alternative explanation.} Independently of the core composition (e.g. icy or rocky with a large water fraction), the question of why these planets begin their evolution with a small primary envelope remains unanswered.

We showed that already a crude parametrisation of the results of planet formation models in combination with detailed atmospheric evolution models driven by thermal contraction and mass loss lead to fit well the observed MR distribution. Our analysis also demonstrates that shaping this distribution cannot be attributed to one specific process, such as atmospheric accretion or escape, but is guided instead by a combination of a wide range of processes, including formation of planetary systems, planetary migration, and interactions between close-in planets and their host stars. Therefore, to improve our understanding of the observed planetary distribution, these processes have to be considered simultaneously in order to use the MR distribution as a diagnostic of planet formation \citep[e.g.][]{venturini2020}. 

\begin{acknowledgements}
     This research has made use of the NASA Exoplanet Archive, which is operated by the California Institute of Technology, under contract with the National Aeronautics and Space Administration under the Exoplanet Exploration Program.
\end{acknowledgements}

%
\bibliographystyle{aa} 
\bibliography{MRrelation.bib} 
%

%
\begin{appendix} 
\section{Systems deviating from the predicted MR distribution}\label{apx::outliers}
\begin{figure*}[h]
   \centering
   \includegraphics[width=0.8\hsize]{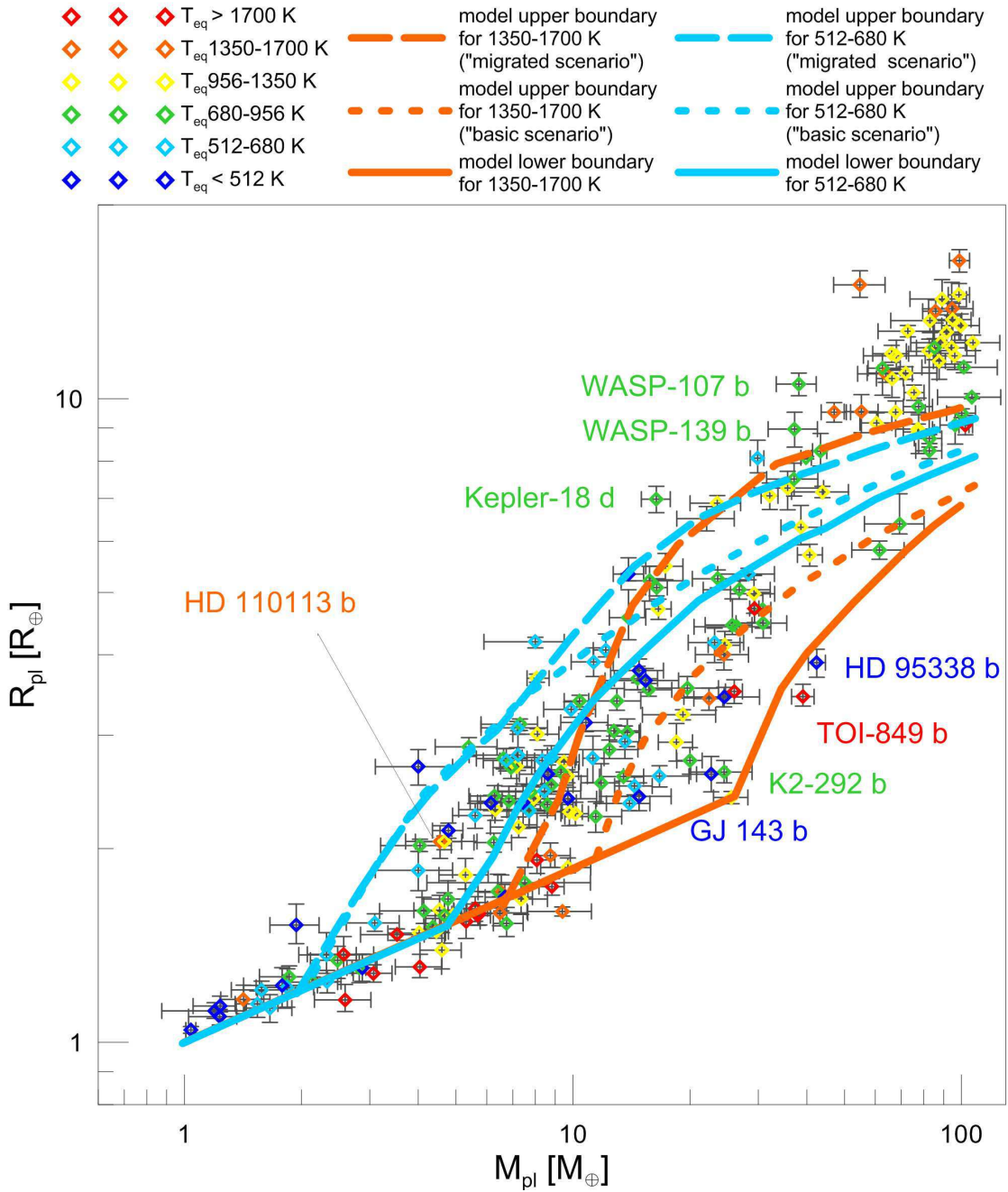} 
   \caption{Same as Figure~\ref{fig::MR_teq}, but highlighting the outliers discussed in Section~\ref{apx::outliers}.}
   \label{fig::MR_teq_apx}
   \end{figure*}

\begin{figure*}[h]
   \centering
   \includegraphics[width=0.8\hsize]{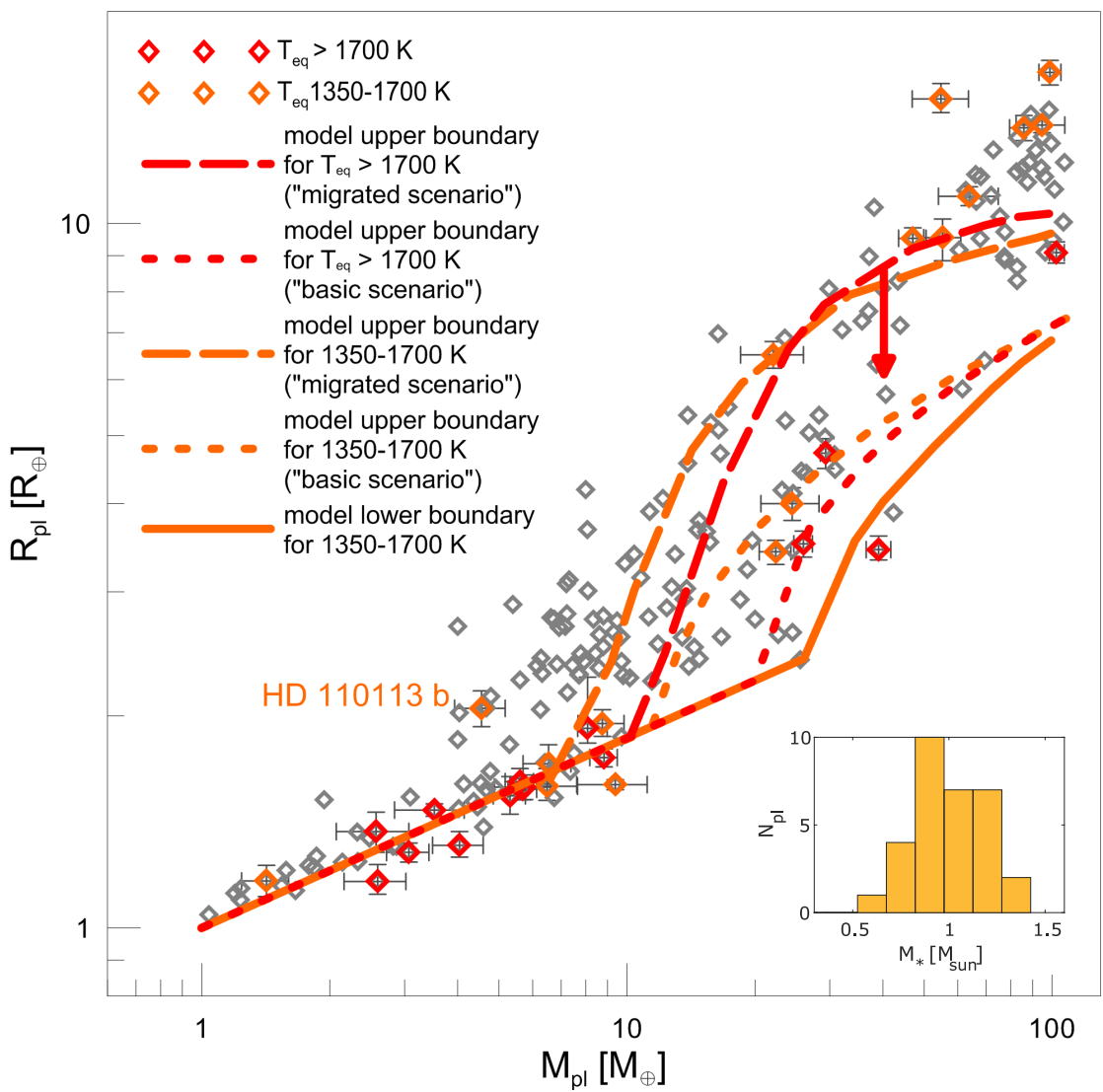} 
   \caption{Same as Figure~\ref{fig::MR_teq_cool}, but for $T_{\rm eq}> 1350$\,K.}
   \label{fig::MR_teq_apx_1350-2500}
   \end{figure*}

\begin{figure*}[h]
   \centering
   \includegraphics[width=0.8\hsize]{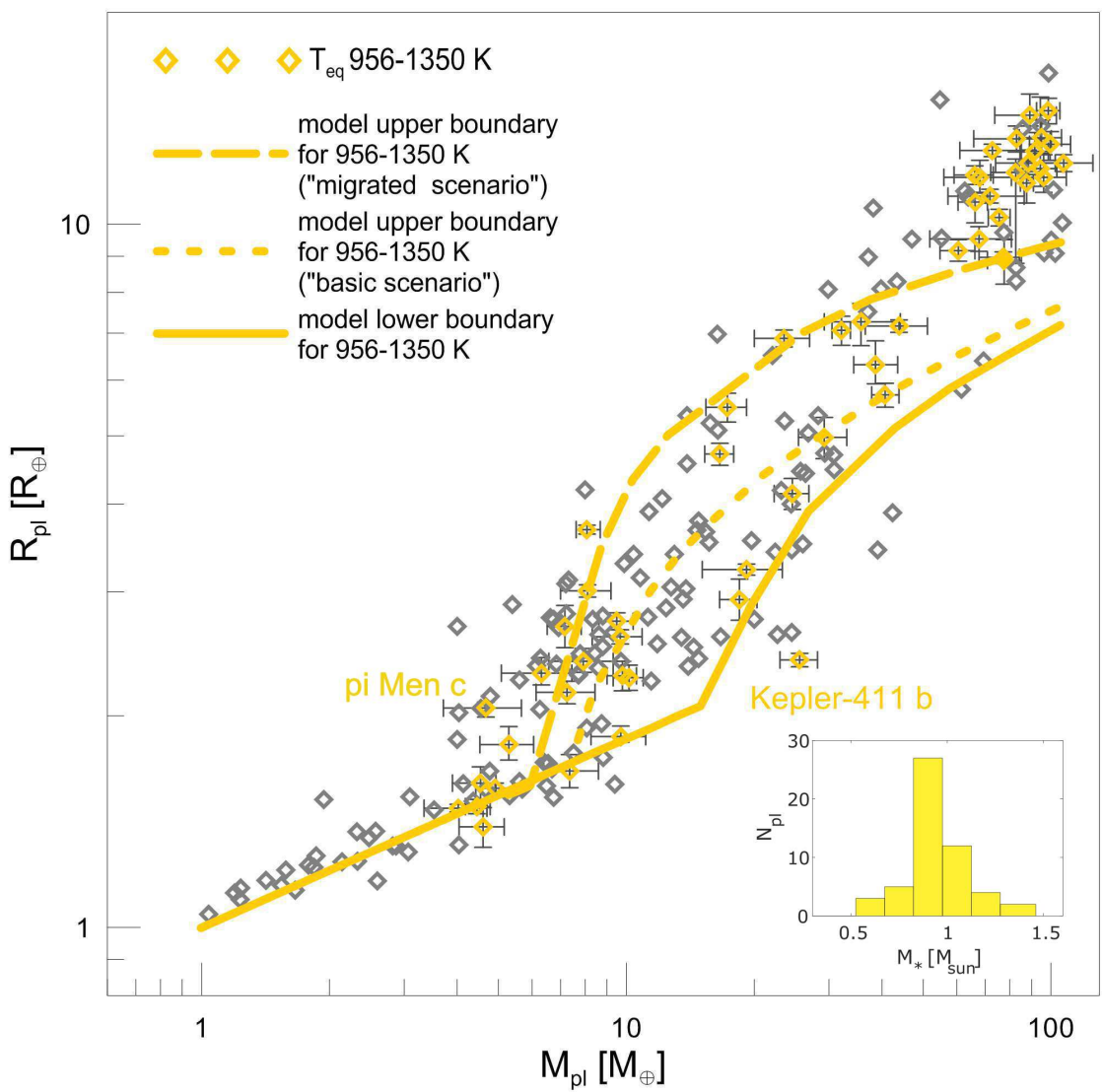} 
   \caption{Same as Figure~\ref{fig::MR_teq_cool}, but for $T_{\rm eq} = 956-1350$\,K.}
   \label{fig::MR_teq_apx_956-1350}
   \end{figure*}

\begin{figure*}[h]
   \centering
   \includegraphics[width=0.8\hsize]{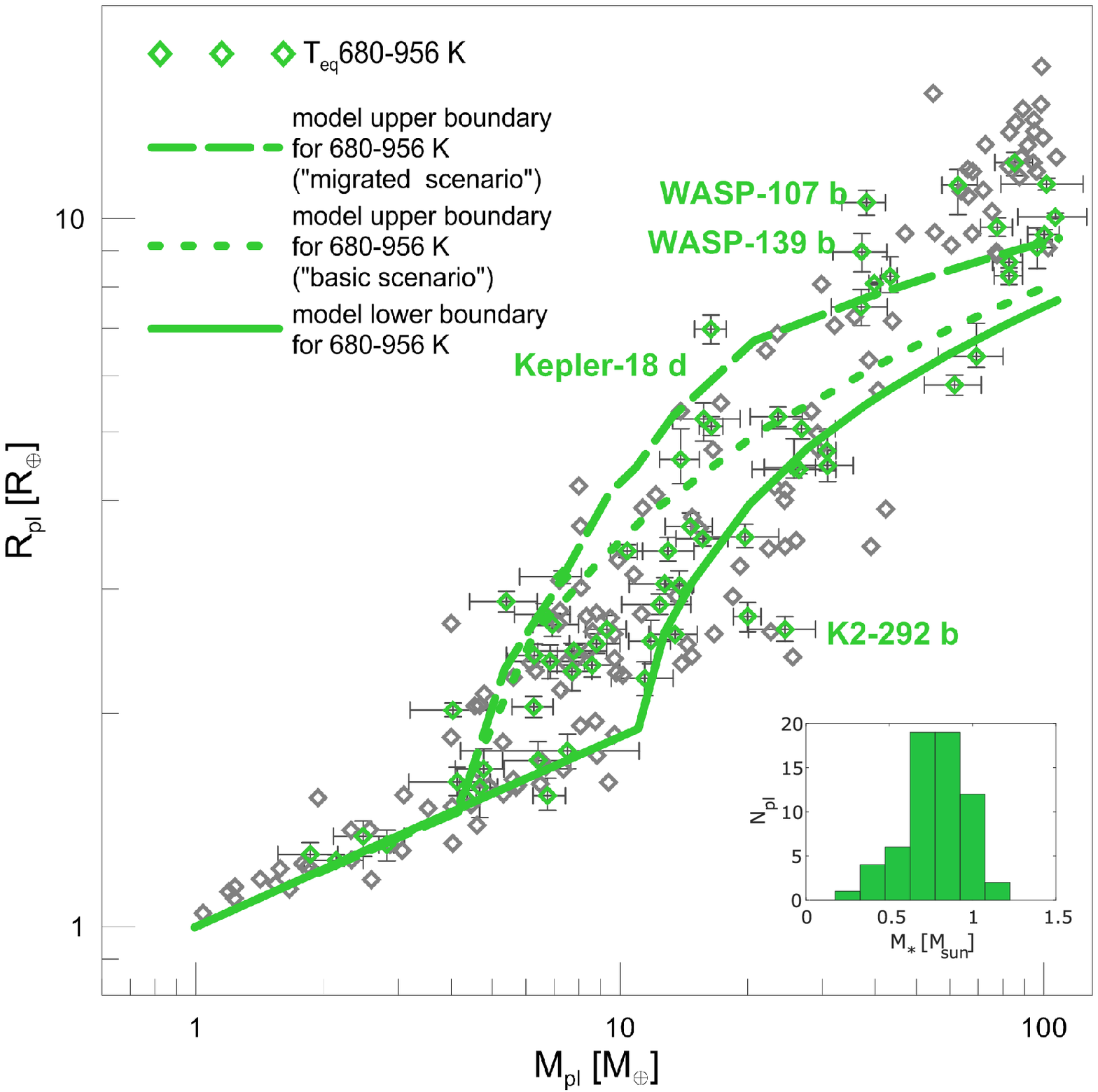} 
   \caption{Same as Figure~\ref{fig::MR_teq_cool}, but for $T_{\rm eq} = 680-956$\,K.}
   \label{fig::MR_teq_apx_680-956}
   \end{figure*}

We thoroughly discuss below the case of TOI-849\,b, while for the other planets we aim mainly at confirming whether they are real outliers or their conflicting position in the MR diagram is the result of specific system parameters, such as the stellar host mass being significantly different from the 1\,$M_{\odot}$ considered in this work. All planets discussed in this section are highlighted in Figure\,\ref{fig::MR_teq_apx}. Figures\,\ref{fig::MR_teq_apx_1350-2500}--\ref{fig::MR_teq_apx_680-956} are analogous to Figure\,\ref{fig::MR_teq_cool}, but show the hotter temperature intervals, namely $T_{\rm eq}>1350$\,K, 956-1359\,K, and 680-956\,K.
%

\subsection{TOI-849\,b}
TOI-849\,b is an outlier compared not only to the modeling predictions, but also (and to a greater extent) to the rest of the observed population. \citet{armstrong2020toi849b} reported a planetary radius of $3.444^{+0.157}_{-0.115}$\,\Rer\ and mass of $39.09^{+2.66}_{-2.55}$\,\Mer, which correspond to an Earth-like density of $5.2_{-0.8}^{+0.7}$\,g/cm$^3$. Therefore, TOI-849\,b appears to be a $\approx$3.5\,\Rer\ bare core for which internal structure modeling concluded that the hydrogen-helium fraction does not exceed $3.9_{-0.9}^{+0.8}$\% of the total planetary mass \citep{armstrong2020toi849b}. This is actually similar to the model planets of similar mass and orbital separation within our ``basic setup'' when assuming the fast rotating host star, which have a hydrogen-helium fraction of 2--3\% at the age of TOI-849\,b ($\sim$7\,Gyr). %

\citet{pezzotti2021toi-849b} studied in detail the possible migration history of TOI-849\,b as a result of stellar dynamical tides, assuming different initial planetary orbits and masses. They concluded that the planet could lie in its current orbit only if it was there already at the time of protoplanetary disk dispersal and the host star evolved as a slow or moderate rotator (the initial stellar rotation rate did not exceed $5\times\Omega_{\odot}$, which is nearly equivalent to the moderate rotator case considered in this work). They also found that in this scenario the atmosphere of TOI-849\,b fully escaped within 30\,Myr independently of the initial atmospheric mass fraction and planetary mass, which is however not consistent with our results as described below.

In our ``basic setup'' scenario, the initial atmospheric mass fraction of TOI-849\,b would be 15\%. However, as mentioned in Section~\ref{sec::model::initials}, such gas-poor formation scenario for a planet at such close orbital separation is rather unlikely, which is confirmed by the absence of compact massive planets at the 0.03\,AU (see the solid and the short-dashed lines in Figure~\ref{fig::MR_teq}). To verify this scenario considering the actual planetary parameters, we run atmospheric evolution assuming that the star evolved as a moderate rotator and employing a stellar mass of 0.929\,$M_{\odot}$, an initial atmospheric mass fraction of 15\%, and an initial planetary mass of 46\,\Mer. In this case, the model predicts total atmospheric loss within 1.9\,Gyr, with a final planetary radius equal to the core radius ($\sim2.7$\,\Rer), which is however considerably smaller than the observed one ($3.444^{+0.157}_{-0.115}$\,\Rer). The same is true also for the case of a slowly rotating star, though the lifetime of the atmosphere extends to almost 5\,Gyr. Therefore, despite TOI-849\,b is located near the lower boundary predicted for planets with comparable temperature, the assumed ``basic setup'' scenario with a compact initial envelope does not appear to be realistic for this specific planet.

We also test a concurrent evolutionary path, namely that of an evaporated gas giant by assuming that TOI-849\,b started its evolution with a more massive hydrogen-dominated atmosphere. Such a scenario is realistic, because at the orbit of TOI-849\,b also heavier planets can be subject to strong atmospheric escape, with mass-loss rates in the early stages of evolution being potentially even higher than those of planets with more compact envelopes \citep{kubyshkina2021mesa}. Therefore, we run a few additional models with $f_{\rm at,0} = 20 - 60\%$, and consequently higher initial masses and always assuming that the present-day mass of TOI-849\,b is nearly equal to that of the core. Starting from this set up, we run the atmospheric evolution considering slow, moderate, and fast rotating host stars.

In the case of the slow rotator, the measured parameters of TOI-849\,b can be reproduced with an initial atmospheric mass fraction of $\sim30-40$\%, corresponding to $M_{\rm pl,0}\sim 51.8-55.8$\,\Mer. At the lower boundary of this interval, the entire atmosphere is lost within 9\,Gyr, which is smaller than the age given by the upper limit inferred for TOI-849. In the case of the moderate rotator, the required initial atmospheric mass increases to $f_{\rm at,0}\sim50-60\%$ ($M_{\rm pl,0}\sim76.6-95$\,\Mer). Finally, we find that for the fast rotating star, the planetary atmosphere is lost in a time shorter than the age of TOI-849\,b (accounting for the uncertainties) for all considered $f_{\rm at,0}$ up to 60\%, leaving behind the bare core with a radius that is at least 0.5\,\Rer\ smaller than the observed one; for larger initial atmospheric mass fractions the atmosphere becomes unstable, because of the too large planetary radius at the time of the disk dispersal.

The atmospheric escape, and hence the evolution of TOI-849\,b depends strongly on the evolution of its host star. The initial atmospheric mass fraction of 50\% and initial mass of 76.6\,\Mer\ allow one reproducing the planet in the case of the moderate rotator (radius of 3.35--3.86\,\Rer\ at $6.7^{+2.9}_{-2.4}$\,Gyr). A planet with the same initial properties, but evolving around a fast rotator would lose its atmosphere within 5\,Gyr and the maximum radius of the planet within the estimated age uncertainty reaches only $\sim$2.9\,\Rer. Instead, considering a slowly rotating star, the same planet ends up with a radius of 4.0--4.6\,\Rer\ within 4.3-9.6\,Gyr. Therefore, at the orbit of TOI-849\,b even a planet born more massive than 70\,\Mer\ experiences intense atmospheric escape throughout its entire life and the final properties are strongly bound to the rotation history of the host star.

To summarise, TOI-849\,b started its evolution most likely as a hot-Saturn with a mass of $\sim50-85$\,\Mer, while the host star has evolved as a slow or moderate rotator. In the case of a moderate rotator, the initial mass of the planet (and corresponding initial atmospheric mass fraction) was close to the limit for which the atmosphere of the young planet is unstable to escape. Furthermore, given that the computed atmospheric mass-loss rates are high throughout the entire evolution (up to about $5\times10^{10}$\,g\,s$^{-1}$ at the present age of the system), it is possible that the planet is at the verge of losing its entire atmosphere.

As mentioned above, these results are somewhat different from those presented by \citet{pezzotti2021toi-849b}. Although both studies consider similar initial planetary masses and end up with the full escape of the atmosphere, the timescales predicted for this to happen are significantly different: order of Gyr compared to about 30\,Myr obtained by \citet{pezzotti2021toi-849b}. Furthermore, in the frame of our models, it is possible to reproduce the present-day parameters of TOI-849\,b employing larger initial masses and atmospheric mass fractions (30--40\% in the case of the slow rotator and 50--60\% for the moderate rotator), because the disruption of the atmosphere due to over-inflation happens at about twice the mass ($\sim$90\,\Mer\ against 46\,\Mer). This is probably due to differences in the modeling approach. Indeed, the three key components of the atmospheric evolution models (planetary structure models relating atmospheric mass fraction and basic parameters; stellar evolution models; atmospheric escape model) differ between the two works and, although the physics underneath the models is qualitatively similar, the differences lead to evident divergences in extreme cases, such as that of TOI-849\,b.
\subsection{HD110113\,b}
%
HD110113\,b is a hot mini-Neptune located within the radius valley and it is expected to host a hydrogen-dominated atmosphere of 0.1--1\% \citep{osborn2021hd110113}. In the MR diagram, the planetary properties ($4.55\pm0.62$\,\Mer; $2.05\pm0.12$\,\Rer) place it near the upper limit predicted for planets having an equilibrium temperature of $\approx$600\,K, that is half of that of the planet of $\sim$1370\,K \citep{osborn2021hd110113}. The neighboring planet in the mass-radius diagram in the same temperature bin, $\pi$\,Men\,c \citep{Huang2018,Gandolfi2018}, has very similar parameters, but lower $T_{\rm eq}$, which is why we focus here on the more extreme one, that is HD110113\,b.

We probed the evolution of the planetary atmosphere assuming initial atmospheric mass fractions ranging between 1.5\% and 70\% (in the latter case, the mass of the atmosphere is $\sim$5.7 times that of the core and an even larger atmospheric mass fraction would lead to an unstable atmosphere), finding that at the position of HD110113\,b the whole atmosphere escapes within 16\,Myr--1.1\,Gyr. Therefore, according to our models, HD110113\,b should not possess any hydrogen-dominated atmosphere, which makes it an actual outlier with respect to the modeling results. 

The small size of the planet makes the core composition an important factor for interpreting the observed planetary radius. As a matter of fact, given the planetary mass, the rocky core should have a radius of 1.45--1.56\,\Rer, which requires some atmosphere to explain the observed radius, but we note that the measured planetary parameters are also consistent with a bare icy core \citep{zeng2021gap_waterEOS}.

An alternative possibility for explaining the radius of HD110113\,b would be the presence of a secondary atmosphere, eventually containing high-altitude hazes leading to a significant radius enhancement \citep{lammer2016,gao_zhang2020puffs}, which is different to the case of hydrogen-dominated atmospheres considered by our models. 

\subsection{WASP-107\,b}
%

WASP-107\,b is about twice Neptune's mass and has a Jupiter-like radius \citep{DaiWinn2017wasp-107b,mochnik2017wasp-107b,Piaulet2021wasp-107b}, which stands out because of its low density. All radii published for WASP-107\,b significantly exceed the radii of planets with similar mass and temperature, as well as the maximum radius predicted by our theoretical models. If one considers the dependence of $R_{\rm pl}$ on equilibrium temperature (Figure~\ref{fig::R-on-Teq}), WASP-107\,b lies at the lower $T_{\rm eq}$ boundary of this group of planets. This suggests that despite the relatively low $T_{\rm eq}$ of $736\pm17$\,K, atmospheric inflation might also be at work on this planet. 

To confirm that WASP-107\,b is an outlier with respect to our theoretical model and belongs to the ``inflated'' planets, we ran a few additional evolutionary models assuming different $f_{\rm at,0}$, the measured system parameters, and a slowly rotating host star (to decrease mass loss), thus aiming at finding the largest possible predicted planetary radius. Assuming an initial atmospheric mass fraction consistent with our ``basic setup'' scenario of $\sim30$\%, we obtain a radius of 5.2--5.5\,\Rer, which is about half of the measured radius and even smaller than our estimate obtained considering a solar-mass star \citep[in agreement with][]{kubyshkina2021mesa}. %

For larger initial atmospheric mass fractions, mass loss throughout the evolution is relatively low and does not exceed $\sim$12\% of the initial planetary mass. However, at the estimated age of the system, the radius of the planet predicted by our models reaches a maximum value of 8.4\,\Rer\ for an initial atmospheric mass fraction of 65\%. For larger initial atmospheric mass fractions, the atmosphere of the young planet becomes unstable even if assuming a very low initial entropy (temperature) of the planetary core. In fact, for the coldest core and $f_{\rm at,0}$, the photospheric radius at an age of 5\,Myr exceeds $\sim$40 times the planetary Roche radius, which leads to fast atmospheric disruption.

We made the above estimates assuming a slowly rotating star ($P_{\rm rot, 150 Myr} = 15$\,days) to derive an upper limit for the planetary radius. However, the rotation period of the host star of $17.5\pm1.4$\,days is atypically short for the stellar age of $8.30\pm4.30$\,Gyr \citep{mochnik2017wasp-107b}. According to the Mors models, this rotation period is compatible with an age of $\sim0.5-1.3$\,Gyr, depending on the specific rotation evolution of the star. The age estimates reported by \citet[$8.3\pm4.3$\,Gyr][]{mochnik2017wasp-107b} and \citet[$3.4\pm0.7$\,Gyr][]{Piaulet2021wasp-107b} are based on the comparison of the stellar parameters with isochrones, which can be considerably model-dependent. However, even in case the system is as young as 0.5\,Gyr, the maximum radius reached within our test models (excluding unstable atmospheres) is 9.7\,\Rer, which is considerably smaller than the measured planetary radius. At the light of these results, we conclude that the large radius of WASP-107\,b is most likely the result of atmospheric inflation, which can not be reproduced within the frame of our models.

In this respect, the atypically fast rotation of the host star might give a hint about the atmospheric inflation mechanism. The impact of both tidal and induction heating is expected to increase with increasing rotation rate of the host star (in the case of induction heating, the fast rotation implies a stronger magnetic field). Due to the orbital inclination of almost $90^{\circ}$ \citep{mochnik2017wasp-107b} and small orbital separation, induction heating might to be particularly effective \citep{kislyakova2018indheat}. Instead, tidal heating requires an eccentric orbit, which is not the case for WASP-107\,b. Furthermore, due to the small mass and relatively low temperature of WASP-107\,b, both vertical heat transport and Ohmic dissipation (the latter requiring a high planetary magnetic field) are unlikely to be effective. This makes induction heating a possible inflation mechanism in this particular case. However, a separate and detailed study is necessary to draw any firmer conclusion.

\subsection{Kepler-18\,d}
%
Kepler-18\,d is a member of a relatively old ($10.0\pm2.3$\,Gyr) planetary system comprising two outer low-density Neptune-mass planets and an inner super-Earth orbiting a solar-mass star \citep{Cochran2011kepler-18,hadden2014kepler-18,morton2016kepler-18,berger2018kepler-18}. Both c and d planets can be classified as super-puffs, and the relatively large mass uncertainty of the inner planet does not rule out the possibility of it hosting a thin hydrogen-dominated atmosphere. The large radius of $6.98\pm0.33$\,\Rer\ of Kepler-18\,d is about one Earth radius above the upper boundary predicted by our theoretical models (see Figure~\ref{fig::MR_teq} and \ref{fig::MR_teq_cool}). However, these considerations are based on the ``default'' parameters listed in the NASA Exoplanet Archive based on \citet{Cochran2011kepler-18}, but later measurements suggest smaller planetary radii of $6.63^{+0.45}_{-0.43}$\,\Rer\ \citep{hadden2014kepler-18}, $6.04^{+0.46}_{-0.40}$\,\Rer\ \citep{morton2016kepler-18}, and $5.198^{+0.222}_{-0.219}$\,\Rer\ \citep{berger2018kepler-18}, and also slightly smaller planetary masses. These more recent values then place Kepler-18\,d at the upper edge of our synthetic population. Despite this, we decided to consider Kepler-18\,d in more detail, as it belongs to the hottest 30\% of planets within the 600--800\,K $T_{\rm eq}$ bin and orbits a solar-mass star, at odds with its location near the upper boundary. To this end, we run additional evolutionary models assuming different initial atmospheric mass fractions.

As expected, the models considering an initial atmospheric mass fraction of 12\%, compatible with the ``basic setup'' scenario, lead to a too-small radius of $\sim3.2$\,\Rer. To reproduce the present-day planetary parameters published by \citet{Cochran2011kepler-18} and assuming a slowly rotating star, one has to assume an initial atmospheric mass fraction of $\sim$60\%, which is about 1.5 times higher than the one assumed for the planetary mass range in the ``migrated'' scenario. The smaller radii given by the other studies lead to initial mass fractions of about 40\%, which fits well our predictions.

Similarly to the case of WASP-107\,b, the rotation period of the host star of about 12\,days  \citep{Cochran2011kepler-18} is atypically short for its age of $10.0\pm2.3$\,Gyr. In fact, the Mors model predicts a gyrochronological age of about 1--2\,Gyr. At this age, the planetary radius published by \citet{Cochran2011kepler-18} would be compatible with that predicted considering an initial atmospheric mass fraction of 40\%.

To summarize, a detailed analysis shows that Kepler-18\,d should not be considered as a real outlier given that its position in the MR diagram relative to the models can be explained by slightly different initial conditions and/or a possible uncertainty on the age (and eventually radius) determination.

\subsection{HD95338\,b}
%
HD95338\,b \citep{diaz2020hd95338} lies below the lower boundary of the total radius spread predicted by the models. The planet is about twice heavier than Neptune, orbits relatively far from its host star ($T_{\rm eq}\simeq 385$\,K), and has a radius of $3.89^{+0.19}_{-0.2}$\,\Rer, which is below the smallest radii predicted for very close-in planets with $T_{\rm eq}\sim 1700$\,K. To reproduce the observations, \citet{diaz2020hd95338} suggested that the atmosphere of HD95338\,b could be composed of a mixture of ammonia, water, and methane with a heavy elements enrichment of about 90\%. For comparison, we considered a planet with parameters similar to those of HD95338\,b starting its evolution with $\sim40$\% of its mass contained in a hydrogen-dominated atmosphere. The evolution models suggest that such a planet would preserve most of the initial envelope until its present age.

We performed atmospheric evolution simulations assuming two different sets of initial parameters: (1) the planet started its evolution with an initial atmospheric mass fraction of 60\% (and consequently higher mass), and (2) the planet started with parameters comparable to the present-day ones with a median mass of 42.4\,\Mer\ and initial atmospheric mass fractions of 10\%, 7\%, and 5\%. For the host star, we considered a mass of 0.83\,$M_{\odot}$ and a rotation evolution equivalent to the cases of moderate and fast rotators considered in the present study, but for the specific stellar mass.

In all cases, the atmospheric mass loss (induced mainly by stellar irradiation) does not affect much the evolution of the planetary atmosphere (radius), which decreases just by less than 0.1\% of the planetary mass. Therefore, a planet that starts with a voluminous atmosphere ends up at the age of HD95338\,b (about 5\,Gyr) with a radius nearly twice larger than the observed one. The three models starting with initial atmospheric mass fractions of 10\%, 7\%, and 5\% lead to radii of 4.3--4.4\,\Rer, 4.01--4.07\,\Rer, and 3.75--3.8\,\Rer, respectively. The latter two values fall into the uncertainty radius range measured for HD95338\,b, and the mass of the atmosphere remains nearly unchanged throughout the evolution in all three cases. Therefore, our models suggest that either this planet was formed with an initial atmospheric mass fraction similar to that observed now or it lost its primordial hydrogen-dominated atmosphere, assuming it had one, possibly as a result of a catastrophic event.

\subsection{K2-292\,b}
%
K2-292\,b is a warm sub-Neptune orbiting at 0.13\,AU around a solar-like star \citep{Luque2019k2-292b}. Similarly to the case of HD95338\,b, the planet lies below the lower boundary of the total radius spread predicted by the models. The planetary mass of $24.5\pm4.0$\,\Mer\ and radius of $2.63^{+0.12_{-0.10}}$\,\Rer\ result in a bulk density of $7.4_{-1.5}^{+1.6}$\,${\rm g\,cm^{-3}}$, suggesting that the planet is most likely rocky and eventually hosting a secondary atmosphere. However, according to our simulations, if the planet would start with a substantial primordial envelope of $f_{\rm at,0}\sim$10--30\%, which is likely for its mass, it would preserve more than about 60\% of it. 

To confirm that the small radius of K2-292\,b is not a consequence of the specific system configuration, we run a few additional evolutionary models assuming the present-day orbital parameters of K2-292\,b, a fast rotating star (to favour mass loss to obtain the minimum possible radius), and a range of different initial atmospheric mass fractions between 1\% and 70\%. For all computed tracks, we find that the total atmospheric mass loss does not exceed $\sim6.2$\% of the planetary mass and that the present-day radius of the planet can only be reproduced assuming an initial atmospheric mass fraction of 1--2\%.

\subsection{GJ\,143\,b}
%
GJ\,143\,b (aka HD21749\,b) is a dense warm ($\sim422$\,K) sub-Neptune orbiting a K-type main sequence star of $0.73\pm0.07$\,$M_{\odot}$. The planetary mass of $22.7_{-1.9}^{+2.2}$\,\Mer\ and radius of $2.61_{-0.16}^{+0.17}$\,\Rer\ lead to a bulk density of $7.02_{+2.28}^{-1.70}$\,${\rm g\,cm^{-3}}$, which is compatible with a purely rocky composition \citep{Dragomir2019gj143b}. However, if starting with a substantial envelope of 25--30\% as for our ``basic setup'' scenario, such a planet would preserve most of its atmosphere and currently have a radius of $\sim$6\,\Rer. Given that the host star is considerably lighter than the solar-mass star considered by our model, we expect that GJ\,143\,b experienced higher atmospheric loss, finally ending up with a smaller radius than the predicted one. However, at the planetary orbit of 0.1915\,AU, corresponding to an equilibrium temperature of $\sim422$\,K, these differences are not expected to be dramatic \citep{kubyshkina2021mesa}.

To quantify the difference and identify what could be the possible initial parameters of GJ\,143\,b, we run atmospheric evolutionary models assuming initial atmospheric mass fractions in the range 1--30\%. We exclude larger initial atmospheric mass fractions, as the case of GJ\,143\,b is very similar to that of K2-292\,b, except for the cooler temperature of GJ\,143\,b, and thus even lower expected atmospheric escape. Assuming an initial atmospheric mass fraction of 30\%, compatible with the ``basic setup'' scenario, we arrive at a current radius larger than $\sim5.35$\,\Rer. This is only slightly smaller than the predicted one obtained considering a solar mass star, confirming that this is an outlier with respect to our modeling results. To reproduce the observed planet parameters in the frame of our evolutionary models, we find that the initial atmospheric mass fraction should not exceed $\sim$1\%.
%
\section{Radius spread of the observed exoplanet distribution}\label{apx::spread}
Here we present the observational radius spread discussed in Section~\ref{sec::compare::overview}. The data underlying Figure~\ref{fig::SPREAD_vs_mass} without optimisation of the intervals are listed in Tables~\ref{apx::tab::allteq} (all temperatures), \ref{apx::tab::tleq1000} ($T_{\rm eq}\leq 1000$~K), and \ref{apx::tab::tgt1000} ($T_{\rm eq}>1000$~K).
\begin{table}\label{apx::tab::allteq}
\caption{Radius spread against planetary mass observed for planets with $T_{\rm eq}$ in 512--1700\,K range.} \label{tab:spread_allpoints}
\begin{tabular}{c|c|c}
\hline
\hline
$M_{\rm pl}$ [\Mer] & $\Delta R_{\rm pl}$ [\Rer] &  $N_{\rm planets}$ \\
   & $R_{\rm pl} < 11$~\Rer/ all $R_{\rm pl}$ & $R_{\rm pl} < 11$~\Rer/ all $R_{\rm pl}$ \\
\hline
  1.06 & -- / --  &  0 / 0 \\ 
  1.17 & -- / --  &  1 / 1 \\ 
  1.29 & -- / --  &  2 / 2 \\ 
  1.42 & $0.076_{-0.095}^{+0.095}$ / $0.076_{-0.095}^{+0.095}$  &  4 / 4 \\ 
  1.57 & $0.134_{-0.105}^{+0.106}$ / $0.134_{-0.105}^{+0.106}$  &  6 / 6 \\ 
  1.73 & $0.134_{-0.105}^{+0.106}$ / $0.134_{-0.105}^{+0.106}$  &  5 / 5 \\ 
  1.91 & $0.134_{-0.103}^{+0.106}$ / $0.134_{-0.103}^{+0.106}$  &  5 / 5 \\ 
  2.10 & $0.150_{-0.134}^{+0.196}$ / $0.150_{-0.134}^{+0.196}$  &  6 / 6 \\ 
  2.32 & $0.130_{-0.113}^{+0.157}$ / $0.130_{-0.113}^{+0.157}$  &  5 / 5 \\ 
  2.55 & $0.293_{-0.069}^{+0.098}$ / $0.293_{-0.069}^{+0.098}$  &  7 / 7 \\ 
  2.82 & $0.291_{-0.097}^{+0.100}$ / $0.291_{-0.097}^{+0.100}$  &  6 / 6 \\ 
  3.10 & $0.228_{-0.109}^{+0.118}$ / $0.228_{-0.109}^{+0.118}$  &  3 / 3 \\ 
  3.42 & $0.717_{-0.106}^{+0.113}$ / $0.717_{-0.106}^{+0.113}$  &  6 / 6 \\ 
  3.77 & $0.572_{-0.090}^{+0.147}$ / $0.572_{-0.090}^{+0.147}$  &  8 / 8 \\ 
  4.16 & $0.661_{-0.149}^{+0.209}$ / $0.661_{-0.149}^{+0.209}$  &  13 / 13 \\ 
  4.58 & $1.488_{-0.186}^{+0.186}$ / $1.488_{-0.186}^{+0.186}$  &  15 / 15 \\ 
  5.05 & $1.488_{-0.186}^{+0.186}$ / $1.488_{-0.186}^{+0.186}$  &  12 / 12 \\ 
  5.57 & $1.488_{-0.186}^{+0.186}$ / $1.488_{-0.186}^{+0.186}$  &  17 / 17 \\ 
  6.14 & $1.590_{-0.160}^{+0.130}$ / $1.590_{-0.160}^{+0.130}$  &  20 / 20 \\ 
  6.77 & $2.660_{-0.180}^{+0.140}$ / $2.660_{-0.180}^{+0.140}$  &  26 / 26 \\ 
  7.47 & $2.660_{-0.180}^{+0.140}$ / $2.660_{-0.180}^{+0.140}$  &  31 / 31 \\ 
  8.23 & $2.593_{-0.116}^{+0.096}$ / $2.593_{-0.116}^{+0.096}$  &  29 / 29 \\ 
  9.07 & $2.593_{-0.116}^{+0.096}$ / $2.593_{-0.116}^{+0.096}$  &  23 / 23 \\ 
 10.00 & $2.473_{-0.126}^{+0.266}$ / $2.473_{-0.126}^{+0.266}$  &  20 / 20 \\ 
 11.03 & $2.473_{-0.126}^{+0.266}$ / $2.473_{-0.126}^{+0.266}$  &  17 / 17 \\ 
 12.16 & $2.328_{-0.469}^{+0.603}$ / $2.328_{-0.469}^{+0.603}$  &  17 / 17 \\ 
 13.40 & $2.973_{-0.488}^{+0.399}$ / $2.973_{-0.488}^{+0.399}$  &  17 / 17 \\ 
 14.77 & $4.630_{-0.390}^{+0.390}$ / $4.630_{-0.390}^{+0.390}$  &  17 / 17 \\ 
 16.29 & $4.630_{-0.390}^{+0.390}$ / $4.630_{-0.390}^{+0.390}$  &  17 / 17 \\ 
 17.95 & $4.388_{-0.428}^{+0.428}$ / $4.388_{-0.428}^{+0.428}$  &  11 / 11 \\ 
 19.79 & $4.388_{-0.387}^{+0.428}$ / $4.388_{-0.387}^{+0.428}$  &  14 / 14 \\ 
 21.82 & $4.481_{-0.244}^{+0.244}$ / $4.481_{-0.244}^{+0.244}$  &  15 / 15 \\ 
 24.05 & $4.481_{-0.244}^{+0.244}$ / $4.481_{-0.244}^{+0.244}$  &  14 / 14 \\ 
 26.52 & $5.679_{-0.463}^{+0.593}$ / $5.679_{-0.463}^{+0.593}$  &  18 / 18 \\ 
 29.23 & $5.679_{-0.463}^{+0.593}$ / $5.679_{-0.463}^{+0.593}$  &  13 / 13 \\ 
 32.22 & $6.056_{-0.656}^{+0.656}$ / $6.056_{-0.656}^{+0.656}$  &  12 / 12 \\ 
 35.52 & $6.056_{-0.656}^{+0.656}$ / $6.056_{-0.656}^{+0.656}$  &  12 / 12 \\ 
 39.16 & $4.819_{-0.650}^{+0.650}$ / $4.819_{-0.650}^{+0.650}$  &  10 / 10 \\ 
 43.17 & $4.819_{-0.650}^{+0.650}$ / $4.819_{-0.650}^{+0.650}$  &  10 / 10 \\ 
 47.59 & $3.833_{-0.582}^{+0.818}$ / $9.309_{-0.879}^{+1.002}$  &  6 / 7 \\ 
 52.47 & $3.730_{-0.548}^{+0.784}$ / $9.206_{-0.846}^{+0.968}$  &  6 / 8 \\ 
 57.84 & $5.109_{-0.515}^{+0.537}$ / $9.206_{-0.846}^{+0.968}$  &  7 / 11 \\ 
 63.77 & $5.142_{-0.437}^{+0.504}$ / $9.206_{-0.846}^{+0.968}$  &  10 / 15 \\ 
 70.30 & $5.142_{-0.437}^{+0.504}$ / $7.407_{-0.750}^{+0.750}$  &  13 / 20 \\ 
 77.49 & $4.572_{-0.974}^{+0.539}$ / $7.896_{-1.226}^{+1.250}$  &  11 / 22 \\ 
 85.43 & $2.667_{-0.471}^{+0.471}$ / $8.093_{-0.885}^{+0.885}$  &  9 / 26 \\ 
 94.18 & $1.759_{-0.280}^{+0.347}$ / $8.093_{-0.885}^{+0.885}$  &  7 / 24 \\ 
103.82 & $0.954_{-0.596}^{+0.723}$ / $7.288_{-1.201}^{+1.261}$  &  3 / 17 \\ 
\hline
\end{tabular}
\end{table}

\begin{table}\label{apx::tab::tleq1000}
\textcolor{dk}{
\caption{Radius spread against planetary mass observed for planets with $T_{\rm eq}\leq 956$\,K.} \label{tab:spread_allpoints_lowTeq}
\begin{tabular}{c|c|c}
\hline
\hline
$M_{\rm pl}$ [\Mer] & $\Delta R_{\rm pl}$ [\Rer] &  $N_{\rm planets}$ \\
   & $R_{\rm pl} < 11$~\Rer/ all $R_{\rm pl}$ & $R_{\rm pl} < 11$~\Rer/ all $R_{\rm pl}$ \\
\hline
  1.07 & -- / --  &  0 / 0 \\ 
  1.19 & -- / --  &  0 / 0 \\ 
  1.31 & -- / --  &  2 / 2 \\ 
  1.45 & $0.076_{-0.095}^{+0.095}$ / $0.076_{-0.095}^{+0.095}$  &  3 / 3 \\ 
  1.61 & $0.134_{-0.105}^{+0.106}$ / $0.134_{-0.105}^{+0.106}$  &  5 / 5 \\ 
  1.78 & $0.134_{-0.103}^{+0.106}$ / $0.134_{-0.103}^{+0.106}$  &  6 / 6 \\ 
  1.97 & $0.237_{-0.143}^{+0.169}$ / $0.237_{-0.143}^{+0.169}$  &  6 / 6 \\ 
  2.18 & $0.150_{-0.134}^{+0.196}$ / $0.150_{-0.134}^{+0.196}$  &  6 / 6 \\ 
  2.41 & $0.128_{-0.111}^{+0.160}$ / $0.128_{-0.111}^{+0.160}$  &  5 / 5 \\ 
  2.67 & $0.291_{-0.097}^{+0.100}$ / $0.291_{-0.097}^{+0.100}$  &  5 / 5 \\ 
  2.95 & $0.228_{-0.109}^{+0.118}$ / $0.228_{-0.109}^{+0.118}$  &  3 / 3 \\ 
  3.27 & $0.545_{-0.193}^{+0.197}$ / $0.545_{-0.193}^{+0.197}$  &  3 / 3 \\ 
  3.62 & $0.511_{-0.090}^{+0.093}$ / $0.511_{-0.090}^{+0.093}$  &  5 / 5 \\ 
  4.00 & $0.511_{-0.090}^{+0.131}$ / $0.511_{-0.090}^{+0.131}$  &  6 / 6 \\ 
  4.43 & $1.367_{-0.143}^{+0.181}$ / $1.367_{-0.143}^{+0.181}$  &  7 / 7 \\ 
  4.91 & $1.367_{-0.143}^{+0.181}$ / $1.367_{-0.143}^{+0.181}$  &  7 / 7 \\ 
  5.43 & $1.305_{-0.230}^{+0.243}$ / $1.305_{-0.230}^{+0.243}$  &  8 / 8 \\ 
  6.01 & $1.590_{-0.160}^{+0.130}$ / $1.590_{-0.160}^{+0.130}$  &  13 / 13 \\ 
  6.65 & $2.660_{-0.180}^{+0.140}$ / $2.660_{-0.180}^{+0.140}$  &  17 / 17 \\ 
  7.36 & $2.660_{-0.180}^{+0.140}$ / $2.660_{-0.180}^{+0.140}$  &  21 / 21 \\ 
  8.15 & $2.660_{-0.180}^{+0.140}$ / $2.660_{-0.180}^{+0.140}$  &  19 / 19 \\ 
  9.02 & $2.421_{-0.191}^{+0.170}$ / $2.421_{-0.191}^{+0.170}$  &  13 / 13 \\ 
  9.98 & $1.828_{-0.223}^{+0.363}$ / $1.828_{-0.223}^{+0.363}$  &  13 / 13 \\ 
 11.05 & $1.828_{-0.223}^{+0.363}$ / $1.828_{-0.223}^{+0.363}$  &  12 / 12 \\ 
 12.23 & $2.328_{-0.473}^{+0.603}$ / $2.328_{-0.473}^{+0.603}$  &  16 / 16 \\ 
 13.53 & $4.738_{-0.453}^{+0.453}$ / $4.738_{-0.453}^{+0.453}$  &  19 / 19 \\ 
 14.98 & $4.630_{-0.390}^{+0.390}$ / $4.630_{-0.390}^{+0.390}$  &  15 / 15 \\ 
 16.58 & $4.630_{-0.390}^{+0.390}$ / $4.630_{-0.390}^{+0.390}$  &  13 / 13 \\ 
 18.35 & $4.388_{-0.428}^{+0.428}$ / $4.388_{-0.428}^{+0.428}$  &  7 / 7 \\ 
 20.31 & $2.658_{-0.268}^{+0.268}$ / $2.658_{-0.268}^{+0.268}$  &  6 / 6 \\ 
 22.48 & $2.620_{-0.280}^{+0.270}$ / $2.620_{-0.280}^{+0.270}$  &  8 / 8 \\ 
 24.88 & $5.450_{-0.520}^{+0.640}$ / $5.450_{-0.520}^{+0.640}$  &  8 / 8 \\ 
 27.54 & $5.450_{-0.520}^{+0.640}$ / $5.450_{-0.520}^{+0.640}$  &  10 / 10 \\ 
 30.48 & $4.546_{-0.622}^{+0.731}$ / $4.546_{-0.622}^{+0.731}$  &  9 / 9 \\ 
 33.73 & $6.056_{-0.656}^{+0.656}$ / $6.056_{-0.656}^{+0.656}$  &  8 / 8 \\ 
 37.34 & $6.056_{-0.656}^{+0.656}$ / $6.056_{-0.656}^{+0.656}$  &  7 / 7 \\ 
 41.32 & $3.036_{-0.866}^{+0.866}$ / $3.036_{-0.866}^{+0.866}$  &  5 / 5 \\ 
 45.74 & $2.439_{-0.462}^{+0.663}$ / $2.439_{-0.462}^{+0.663}$  &  3 / 3 \\ 
 50.62 & -- / --  &  2 / 2 \\ 
 56.03 & -- / --  &  1 / 2 \\ 
 62.01 & -- / $5.333_{-1.221}^{+0.773}$  &  2 / 3 \\ 
 68.64 & $3.898_{-0.470}^{+0.493}$ / $5.333_{-1.221}^{+0.773}$  &  5 / 6 \\ 
 75.97 & $3.328_{-1.007}^{+0.528}$ / $5.610_{-1.227}^{+0.625}$  &  4 / 6 \\ 
 84.08 & $3.328_{-1.007}^{+0.528}$ / $5.610_{-1.227}^{+0.625}$  &  6 / 8 \\ 
 93.06 & $1.759_{-0.280}^{+0.347}$ / $3.705_{-0.724}^{+0.624}$  &  6 / 8 \\ 
103.00 & $0.954_{-0.596}^{+0.723}$ / $2.900_{-1.040}^{+1.000}$  &  3 / 5 \\ 
\hline
\end{tabular}
}
\end{table}

\begin{table}\label{apx::tab::tgt1000}
\textcolor{dk}{
\caption{Radius spread against planetary mass observed for planets with $T_{\rm eq}> 956$\,K.} \label{tab:spread_allpoints_highTeq}
\begin{tabular}{c|c|c}
\hline
\hline
$M_{\rm pl}$ [\Mer] & $\Delta R_{\rm pl}$ [\Rer] &  $N_{\rm planets}$ \\
   & $R_{\rm pl} < 11$~\Rer/ all $R_{\rm pl}$ & $R_{\rm pl} < 11$~\Rer/ all $R_{\rm pl}$ \\
\hline
  1.07 & -- / --  &  0 / 0 \\ 
  1.19 & -- / --  &  1 / 1 \\ 
  1.31 & -- / --  &  1 / 1 \\ 
  1.45 & -- / --  &  1 / 1 \\ 
  1.61 & -- / --  &  1 / 1 \\ 
  1.78 & -- / --  &  0 / 0 \\ 
  1.97 & -- / --  &  0 / 0 \\ 
  2.18 & -- / --  &  1 / 1 \\ 
  2.41 & -- / --  &  1 / 1 \\ 
  2.67 & -- / --  &  1 / 1 \\ 
  2.95 & -- / --  &  1 / 1 \\ 
  3.27 & -- / --  &  0 / 0 \\ 
  3.62 & -- / --  &  1 / 1 \\ 
  4.00 & $0.661_{-0.149}^{+0.209}$ / $0.661_{-0.149}^{+0.209}$  &  6 / 6 \\ 
  4.43 & $0.661_{-0.149}^{+0.209}$ / $0.661_{-0.149}^{+0.209}$  &  8 / 8 \\ 
  4.91 & $0.661_{-0.149}^{+0.209}$ / $0.661_{-0.149}^{+0.209}$  &  8 / 8 \\ 
  5.43 & $0.909_{-0.177}^{+0.210}$ / $0.909_{-0.177}^{+0.210}$  &  10 / 10 \\ 
  6.01 & $1.103_{-0.207}^{+0.250}$ / $1.103_{-0.207}^{+0.250}$  &  8 / 8 \\ 
  6.65 & $2.090_{-0.149}^{+0.126}$ / $2.090_{-0.149}^{+0.126}$  &  9 / 9 \\ 
  7.36 & $2.090_{-0.149}^{+0.126}$ / $2.090_{-0.149}^{+0.126}$  &  10 / 10 \\ 
  8.15 & $2.082_{-0.080}^{+0.080}$ / $2.082_{-0.080}^{+0.080}$  &  12 / 12 \\ 
  9.02 & $2.082_{-0.080}^{+0.080}$ / $2.082_{-0.080}^{+0.080}$  &  10 / 10 \\ 
  9.98 & $1.129_{-0.101}^{+0.101}$ / $1.129_{-0.101}^{+0.101}$  &  7 / 7 \\ 
 11.05 & $1.129_{-0.101}^{+0.101}$ / $1.129_{-0.101}^{+0.101}$  &  6 / 6 \\ 
 12.23 & -- / --  &  1 / 1 \\ 
 13.53 & -- / --  &  0 / 0 \\ 
 14.98 & -- / --  &  2 / 2 \\ 
 16.58 & $2.564_{-0.462}^{+0.451}$ / $2.564_{-0.462}^{+0.451}$  &  4 / 4 \\ 
 18.35 & $3.584_{-0.482}^{+0.481}$ / $3.584_{-0.482}^{+0.481}$  &  6 / 6 \\ 
 20.31 & $3.956_{-0.393}^{+0.382}$ / $3.956_{-0.393}^{+0.382}$  &  9 / 9 \\ 
 22.48 & $4.481_{-0.244}^{+0.244}$ / $4.481_{-0.244}^{+0.244}$  &  8 / 8 \\ 
 24.88 & $4.481_{-0.244}^{+0.244}$ / $4.481_{-0.244}^{+0.244}$  &  7 / 7 \\ 
 27.54 & $4.661_{-0.389}^{+0.389}$ / $4.661_{-0.389}^{+0.389}$  &  6 / 6 \\ 
 30.48 & $4.862_{-0.590}^{+0.502}$ / $4.862_{-0.590}^{+0.502}$  &  4 / 4 \\ 
 33.73 & $2.286_{-0.873}^{+0.785}$ / $2.286_{-0.873}^{+0.785}$  &  5 / 5 \\ 
 37.34 & $1.546_{-0.470}^{+0.673}$ / $1.546_{-0.470}^{+0.673}$  &  5 / 5 \\ 
 41.32 & $3.811_{-0.560}^{+0.560}$ / $3.811_{-0.560}^{+0.560}$  &  5 / 5 \\ 
 45.74 & $3.833_{-0.582}^{+0.818}$ / $9.309_{-0.879}^{+1.002}$  &  5 / 6 \\ 
 50.62 & $2.387_{-0.504}^{+0.740}$ / $7.862_{-0.801}^{+0.924}$  &  4 / 5 \\ 
 56.03 & $1.759_{-0.665}^{+0.662}$ / $5.856_{-0.996}^{+1.093}$  &  6 / 9 \\ 
 62.01 & $1.792_{-0.587}^{+0.629}$ / $5.856_{-0.996}^{+1.093}$  &  7 / 11 \\ 
 68.64 & $2.084_{-0.504}^{+0.976}$ / $4.349_{-0.817}^{+1.222}$  &  8 / 13 \\ 
 75.97 & $2.084_{-0.504}^{+0.976}$ / $5.407_{-0.755}^{+1.688}$  &  7 / 17 \\ 
 84.08 & $2.084_{-0.504}^{+0.909}$ / $7.510_{-0.918}^{+1.323}$  &  4 / 19 \\ 
 93.06 & -- / $7.510_{-0.918}^{+1.323}$  &  2 / 17 \\ 
 103.00 & -- / $4.944_{-1.513}^{+1.390}$  &  0 / 13 \\ 
\hline
\end{tabular}
}
\end{table}
%
\end{appendix}
\end{document}